\documentclass[preprint,showpacs,superscriptaddress]{revtex4}
\usepackage[pdftex]{graphicx}
\usepackage{amssymb,amsfonts,amsmath,verbatim}
\usepackage{epstopdf}
\usepackage{setspace}

\begin{document}
\title{Non-Markovian full counting statistics in quantum dot molecules}
\author{Hai-Bin Xue}
\email{xuehaibin@tyut.edu.cn}
\affiliation{College of Physics and Optoelectronics, Taiyuan University of Technology,
Taiyuan 030024, China}
\author{Hu-Jun Jiao}
\affiliation{Department of Physics, Shanxi University, Taiyuan 030006,
China}
\author{Jiu-Qing Liang}
\affiliation{Institute of Theoretical Physics, Shanxi University, Taiyuan 030006,
China}
\author{Wu-Ming Liu}
\email{wliu@iphy.ac.cn}
\affiliation{Beijing National Laboratory for Condensed Matter Physics, Institute of
Physics, Chinese Academy of Sciences, Beijing 100190, China}

\date{\today}

\begin{abstract}
Full counting statistics of electron transport is a powerful diagnostic
tool for probing the nature of quantum transport beyond what is obtainable
from the average current or conductance measurement alone. In particular,
the non-Markovian dynamics of quantum dot molecule plays an important
role in the nonequilibrium electron tunneling processes. It is thus necessary
to understand the non-Markovian full counting statistics in a
quantum dot molecule. Here we study the non-Markovian full counting
statistics in two typical quantum dot molecules, namely, serially coupled and side-coupled
double quantum dots with high quantum coherence in a certain
parameter regime. We demonstrate that the non-Markovian effect manifests
itself through the quantum coherence of the quantum dot molecule system,
and has a significant impact on the full counting statistics in the
high quantum-coherent quantum dot molecule system, which depends on
the coupling of the quantum dot molecule system with the source and
drain electrodes. The results indicated that the influence of the
non-Markovian effect on the full counting statistics of electron
transport, which should be considered in a high quantum-coherent quantum dot molecule
system, can provide a better understanding of electron transport through quantum dot molecules.
\end{abstract}
\maketitle

\section*{Introduction}

Full counting statistics\cite{Levitov} (FCS) of electron transport
through mesoscopic system has attracted considerable attention both
experimentally and theoretically because it can provide a deeper insight into
the nature of electron transport mechanisms, which cannot be obtained from
the average current \cite%
{Blanter,Nazarov,Gustavsson,Fujisawa,Flindt,Fricke01,Ubbelohde,Fricke02,Maisi}%
. For instance, the shot noise measurements can be used to probe the
dynamical in an open double quantum dots (QDs) \cite{Aguado},
the coherent coupling between serially coupled QDs \cite{Kieblich},
the evolution of the Kondo effect in a QD \cite{Yamauchi}, and
the conduction channels of quantum conductors \cite{Vardimon}. In particular, shot noise characteristics can provide information about
the feature of the pseudospin Kondo effect in a laterally coupled double QDs
\cite{Kubo}, the spin accumulations in a electron reservoir \cite{Meair},
and the charge fractionalization in the $\nu =2$ quantum Hall edge \cite%
{MilletarI}. In addition, the degree of entanglement of two electrons in the
double QDs \cite{Bodoky}, the dephasing rate in a closed QD \cite{Dubrovin},
the internal level structure of single molecule magnet \cite%
{Xuejap10,Xuepla11} can be characterized by the
super-Poissonian shot noise .

On the other hand, the quantum coherence in coupled QD system, which is
characterized by the off-diagonal elements of the reduced density matrix of
the QD system within the framework of the density matrix theory\cite{Blum},
plays an important role in the electron tunneling processes and has a
significant influence on electron transport \cite%
{Gurvitz,Braun,Wunsch,Djuric01,Djuric02,Harbola,Pedersen,Begemann,Darau,Schultz,Schaller}%
. In particular, theoretical studies have demonstrated that the high-order
cumulants, e.g., the shot noise, the skewness, are more sensitive to the
quantum coherence than the average current in the different types of QD
systems \cite{Kieblich,Kieblich01,Xueaop,Welack,Fang01,Fang02} and the
quantum coherence information in a side-coupled double QD system can be
extracted from the high-order current cumulants \cite{Xueaop}. In fact, the
non-Markovian dynamics of the QD system also plays an important role in the
non-equilibrium electron tunneling processes. However, the above studies on
current noise or FCS were mainly based on the different types of Markovian
master equations. Although the influence of non-Markovian effect on the
long-time limit of the FCS in the QD systems has received some attention
\cite%
{Schaller,Braggio01,Flindt01,Zedler01,Emary01,Flindt02,Marcos,Emary02,Zedler02}%
, how the non-Markovian effect affects the FCS is still an open issue,
especially the influence of the interplay between the quantum coherence and
non-Markovian effect on the long-time limit of the FCS has not yet been
revealed.

The aim of this report is thus to derive a non-Markovian FCS formalism
based on the exact time-convolutionless (TCL) master equation and
study the influences of the quantum coherence and non-Markovian effect on
the FCS in QD molecule systems. It is demonstrated that the non-Markovian
effect manifests itself through the quantum coherence of the considered QD
molecule system, and has a significant impact on the FCS in the
high quantum-coherent QD molecule system, which depends on the coupling of the
considered QD molecule system with the incident and outgoing electrodes. Consequently,
it is necessary to consider the influence of the non-Markovian effect on the full
counting statistics of electron transport in a high quantum-coherent single-molecule system.

\section*{Results}

We now study the influences of the quantum coherence and non-Markovian effect
on the FCS of electronic transport through the QD molecule system. In order to facilitate
discussions effectively, we consider three typical QD systems, namely, single QD
without quantum coherence, serially coupled double QDs and side-coupled
double QDs with high quantum coherence in a certain parameter regime (see Fig. 1). In addition, we
assume the bias voltage ($\mu _{L}=-\mu _{R}=V_{b}/2$) is symmetrically
entirely dropped at the QD-electrode tunnel junctions, which implies that
the levels of the QDs are independent of the applied bias voltage even if
the couplings are not symmetric, and choose meV as the unit of energy which
corresponds to a typical experimental situation\cite{Elzerman}.

\subsection*{Single quantum dot without quantum coherence}

In this subsection, we consider a single QD weakly coupled to two
ferromagnetic electrodes. The Hamiltonian of the considered
system is described by the $H_{total}=H_{dot}+H_{leads}+H_{T}$. The QD
Hamiltonian $H_{dot}$ is given by
\begin{equation}
H_{\text{dot,1}}=\sum_{\sigma }\varepsilon_{\sigma } d_{\sigma }^{\dag }d_{\sigma
}+Ud_{\uparrow }^{\dag }d_{\uparrow }d_{\downarrow }^{\dag }d_{\downarrow },
\label{model1}
\end{equation}%
where $d_{\sigma }^{\dag }$ ($d_{\sigma }$) creates (annihilates) an
electron with spin $\sigma $ and on-site energy $\varepsilon_{\sigma } $ (which can be
tuned by a gate voltage $V_{g}$) in this QD system. $U$ is the intradot
Coulomb interaction between two electrons in the QD system.

The relaxation in the two ferromagnetic electrodes is assumed to be
sufficiently fast, so that their electron distributions can be described by
equilibrium Fermi functions. The two electrodes are thus modeled as
non-interacting Fermi gases and the corresponding Hamiltonians can be
expressed as
\begin{equation}
H_{\text{Leads,1}}=\sum_{\alpha \mathbf{k}s}\varepsilon _{\alpha \mathbf{k}%
}a_{\alpha \mathbf{k}s}^{\dag }a_{\alpha \mathbf{k}s},  \label{Leads1}
\end{equation}%
where $a_{\alpha \mathbf{k}s}^{\dag }$ ($a_{\alpha \mathbf{k}s}$) creates
(annihilates) an electron with energy $\varepsilon _{\alpha \mathbf{k}}$,
spin $s$ and momentum $\mathbf{k}$ in $\alpha $ ($\alpha =L,R$) electrode,
and $s=+\left( -\right) $ denotes the majority (minority) spin states with
the density of states $g_{\alpha ,s}$. The polarization vectors $\mathbf{p}%
_{L}$ (left lead) and $\mathbf{p}_{R}$ (right lead) are parallel to each
other, and their magnitudes are characterized by $p_{\alpha }=\left\vert
\mathbf{p}_{\alpha }\right\vert =\left( g_{\alpha ,+}-g_{\alpha ,-}\right)
/\left( g_{\alpha ,+}+g_{\alpha ,-}\right) $. The tunneling between the QD
and the electrodes is described by
\begin{equation}
H_{\text{T,1}}=t_{Lk+}a_{Lk+}^{\dag }d_{\uparrow }+t_{Rk+}a_{Rk+}^{\dag
}d_{\uparrow }+t_{Lk-}a_{Lk-}^{\dag }d_{\downarrow }+t_{Rk-}a_{Rk-}^{\dag
}d_{\downarrow }+\text{H.c.},  \label{tunneling1}
\end{equation}%
where spin-up $\uparrow $ and spin-down $\downarrow $ are defined to be the
majority spin and minority spin of the ferromagnet, respectively.

The QD-electrode coupling is assumed to be sufficiently weak, thus, the
sequential tunneling is dominant and can be well described by the quantum
master equation of reduced density matrix spanned by the eigenstates of the
QD. The particle-number-resolved TCL quantum master equation for the reduced
density matrix of the considered single QD is given by
\begin{eqnarray}
&&\left. \dot{\rho}^{\left( n\right) }\left( t\right) \right\vert _{\text{%
dot,1}}  \notag \\
&=&-i\mathcal{L}\rho ^{(n)}-\sum_{\alpha \sigma }\left[ A_{\alpha \sigma
}^{(+)}d_{\sigma }^{\dag }+d_{\sigma }A_{\alpha \sigma }^{(-)}\rho
^{(n)}-A_{L\sigma }^{(-)}\rho ^{(n)}d_{\sigma }^{\dag }\right.   \notag \\
&&\left. -A_{R\sigma }^{(-)}\rho ^{(n-1)}d_{\sigma }^{\dag }-d_{\sigma
}^{\dag }\rho ^{(n)}A_{L\sigma }^{(+)}-d_{\sigma }^{\dag }\rho
^{(n+1)}A_{R\sigma }^{(+)}+\text{H.c.}\right] ,  \label{Master1}
\end{eqnarray}%
For more details, see Methods section. Here, the complete basis $\left\{ \left\vert 0,0\right\rangle \right. $, $%
\left\vert \uparrow ,0\right\rangle $, $\left\vert \downarrow
,0\right\rangle $, $\left. \left\vert \uparrow ,\downarrow \right\rangle
\right\} $ is chosen to describe the electronic states of this single QD
system, and the single QD system parameters are chosen as $\epsilon
_{\uparrow }=\epsilon _{\downarrow }=1$, $U=5$, $p=0.9$ and $k_{B}T=0.04$.

Figure 2 shows the first four current cumulants as a function of the bias
voltage for different ratios $\Gamma _{L}/\Gamma _{R}$ describing the
left-right asymmetry of the QD-electrode coupling. We found that the
non-Markovian effect has no influence on the current noise behaviors of the
single QD considered here, see Fig. 2. Scrutinizing Eq. (\ref{Master1}), it is found
that for the non-Markovian case the elements of the reduced density
matrix are equivalent to that for the Markovian case because there are not
the off-diagonal elements of the reduced density matrix. Thus, the equations of
motion of the four elements of the reduced density matrix can be expressed as
\begin{align}
&  \left\langle 0,0\right\vert \dot{\rho}_{S}\left(  t\right)  \left\vert
0,0\right\rangle \nonumber\\
&  =-\left[  \Gamma_{L\uparrow}f_{L,+}\left(  \epsilon_{\uparrow}\right)
+\Gamma_{R\uparrow}f_{R,+}\left(  \epsilon_{\uparrow}\right)  \right]
\left\langle 0,0\right\vert \rho_{S}\left(  t\right)  \left\vert
0,0\right\rangle \nonumber\\
&  -\left[  \Gamma_{L\downarrow}f_{L,+}\left(  \epsilon_{\downarrow}\right)
+\Gamma_{R\downarrow}f_{R,+}\left(  \epsilon_{\downarrow}\right)  \right]
\left\langle 0,0\right\vert \rho_{S}\left(  t\right)  \left\vert
0,0\right\rangle \nonumber\\
&  +\left[  \Gamma_{L\uparrow}f_{L,-}\left(  \epsilon_{\uparrow}\right)
+\Gamma_{R\uparrow}f_{R,-}\left(  \epsilon_{\uparrow}\right)  e^{i\chi
}\right]  \left\langle 0,\uparrow\right\vert \rho_{S}\left(  t\right)
\left\vert \uparrow,0\right\rangle \nonumber\\
&  +\left[  \Gamma_{L\downarrow}f_{L,-}\left(  \epsilon_{\downarrow}\right)
+\Gamma_{R\downarrow}f_{R,-}\left(  \epsilon_{\downarrow}\right)  e^{i\chi
}\right]  \left\langle \downarrow,0\right\vert \rho_{S}\left(  t\right)
\left\vert 0,\downarrow\right\rangle ,  \label{zero1}%
\end{align}
\begin{align}
&  \left\langle 0,\uparrow\right\vert \dot{\rho}_{S}\left(  t\right)
\left\vert \uparrow,0\right\rangle \nonumber\\
&  =\left[  \Gamma_{L\uparrow}f_{L,+}\left(  \epsilon_{\uparrow}\right)
+\Gamma_{R\uparrow}f_{R,+}\left(  \epsilon_{\uparrow}\right)  e^{-i\chi
}\right]  \left\langle 0,0\right\vert \rho_{S}\left(  t\right)  \left\vert
0,0\right\rangle \nonumber\\
&  -\left[  \Gamma_{L\uparrow}f_{L,-}\left(  \epsilon_{\uparrow}\right)
+\Gamma_{R\uparrow}f_{R,-}\left(  \epsilon_{\uparrow}\right)  \right]
\left\langle 0,\uparrow\right\vert \rho_{S}\left(  t\right)  \left\vert
\uparrow,0\right\rangle \nonumber\\
&  -\left[  \Gamma_{L\downarrow}f_{L,+}\left(  \epsilon_{\uparrow\downarrow
}-\epsilon_{\uparrow}\right)  +\Gamma_{R\downarrow}f_{R,+}\left(
\epsilon_{\uparrow\downarrow}-\epsilon_{\uparrow}\right)  \right]
\left\langle 0,\uparrow\right\vert \rho_{S}\left(  t\right)  \left\vert
\uparrow,0\right\rangle \nonumber\\
&  +\left[  \Gamma_{L\downarrow}f_{L,-}\left(  \epsilon_{\uparrow\downarrow
}-\epsilon_{\uparrow}\right)  +\Gamma_{R\downarrow}f_{R,-}\left(
\epsilon_{\uparrow\downarrow}-\epsilon_{\uparrow}\right)  e^{i\chi}\right]
\left\langle \downarrow,\uparrow\right\vert \rho_{S}\left(  t\right)
\left\vert \uparrow,\downarrow\right\rangle , \label{oneup1}%
\end{align}
\begin{align}
&  \left\langle \downarrow,0\right\vert \dot{\rho}_{S}\left(  t\right)
\left\vert 0,\downarrow\right\rangle \nonumber\\
&  =\left[  \Gamma_{L\downarrow}f_{L,+}\left(  \epsilon_{\downarrow}\right)
+\Gamma_{R\downarrow}f_{R,+}\left(  \epsilon_{\downarrow}\right)  e^{-i\chi
}\right]  \left\langle 0,0\right\vert \rho_{S}\left(  t\right)  \left\vert
0,0\right\rangle \nonumber\\
&  -\left[  \Gamma_{L\downarrow}f_{L,-}\left(  \epsilon_{\downarrow}\right)
+\Gamma_{R\downarrow}f_{R,-}\left(  \epsilon_{\downarrow}\right)  \right]
\left\langle \downarrow,0\right\vert \rho_{S}\left(  t\right)  \left\vert
0,\downarrow\right\rangle \nonumber\\
&  -\left[  \Gamma_{L\uparrow}f_{L,+}\left(  \epsilon_{\uparrow\downarrow
}-\epsilon_{\downarrow}\right)  +\Gamma_{R\uparrow}f_{R,+}\left(
\epsilon_{\uparrow\downarrow}-\epsilon_{\downarrow}\right)  \right]
\left\langle \downarrow,0\right\vert \rho_{S}\left(  t\right)  \left\vert
0,\downarrow\right\rangle \nonumber\\
&  +\left[  \Gamma_{L\uparrow}f_{L,-}\left(  \epsilon_{\uparrow\downarrow
}-\epsilon_{\downarrow}\right)  +\Gamma_{R\uparrow}f_{R,-}\left(
\epsilon_{\uparrow\downarrow}-\epsilon_{\downarrow}\right)  e^{i\chi}\right]
\left\langle \downarrow,\uparrow\right\vert \rho_{S}\left(  t\right)
\left\vert \uparrow,\downarrow\right\rangle , \label{onedown1}%
\end{align}
\begin{align}
&  \left\langle \downarrow,\uparrow\right\vert \dot{\rho}_{S}\left(  t\right)
\left\vert \uparrow,\downarrow\right\rangle \nonumber\\
&  =\left[  \Gamma_{L\downarrow}f_{L,+}\left(  \epsilon_{\uparrow\downarrow
}-\epsilon_{\uparrow}\right)  +\Gamma_{R\downarrow}f_{R,+}\left(
\epsilon_{\uparrow\downarrow}-\epsilon_{\uparrow}\right)  e^{-i\chi}\right]
\left\langle 0,\uparrow\right\vert \rho_{S}\left(  t\right)  \left\vert
\uparrow,0\right\rangle \nonumber\\
&  +\left[  \Gamma_{L\uparrow}f_{L,+}\left(  \epsilon_{\uparrow\downarrow
}-\epsilon_{\downarrow}\right)  +\Gamma_{R\uparrow}f_{R,+}\left(
\epsilon_{\uparrow\downarrow}-\epsilon_{\downarrow}\right)  e^{-i\chi}\right]
\left\langle \downarrow,0\right\vert \rho_{S}\left(  t\right)  \left\vert
0,\downarrow\right\rangle \nonumber\\
&  -\left[  \Gamma_{L\uparrow}f_{L,-}\left(  \epsilon_{\uparrow\downarrow
}-\epsilon_{\downarrow}\right)  +\Gamma_{R\uparrow}f_{R,-}\left(
\epsilon_{\uparrow\downarrow}-\epsilon_{\downarrow}\right)  \right]
\left\langle \downarrow,\uparrow\right\vert \rho_{S}\left(  t\right)
\left\vert \uparrow,\downarrow\right\rangle \nonumber\\
&  -\left[  \Gamma_{L\downarrow}f_{L,-}\left(  \epsilon_{\uparrow\downarrow
}-\epsilon_{\uparrow}\right)  +\Gamma_{R\downarrow}f_{R,-}\left(
\epsilon_{\uparrow\downarrow}-\epsilon_{\uparrow}\right)  \right]
\left\langle \downarrow,\uparrow\right\vert \rho_{S}\left(  t\right)
\left\vert \uparrow,\downarrow\right\rangle . \label{two1}%
\end{align}
Here, $f_{\alpha,+}$ is the Fermi function of the electrode $\alpha$, and
$f_{\alpha,-}=1-f_{\alpha,+}$. The detailed procedure for calculation of the equation of
motion of a reduced density matrix, see Methods section. Within the
framework of the density matrix theory, the off-diagonal elements of the
reduced density matrix characterize the quantum coherence of the considered
QD system. Thus, the influence of the non-Markovian effect on the FCS
may be associated with the quantum coherence of the considered QD system. In
order to confirm this conclusion, we take serially coupled and
side-coupled double QDs for illustration in the following two subsection.

\subsection*{Serially coupled double quantum dots with high quantum coherence}

We now consider two serially coupled double QDs weakly connected to two
metallic electrodes, see Fig. 1(a). For the sake of simplicity, the spin degree of freedom
has not been considered. The double-QD is described by a spinless Hamiltonian
\begin{equation}
H_{\text{dot,2}}=\epsilon _{1}d_{1}^{\dag }d_{1}+\epsilon _{2}d_{2}^{\dag
}d_{2}+U\hat{n}_{1}\hat{n}_{2}-J\,\left( d_{1}^{\dag }d_{2}+d_{2}^{\dag
}d_{1}\right) ,  \label{model2}
\end{equation}%
where $d_{i}^{\dag }$ ($d_{i}$) creates (annihilates) an electron with
energy $\varepsilon _{i}$ (which can be tuned by a gate voltage $V_{g}$) in $%
i$th QD. $U$ is the interdot Coulomb repulsion between two electrons in the
double QD system, where we consider the intradot Coulomb interaction $U\rightarrow
\infty ,$ so that the double-electron occupation in the same QD is
prohibited. The last term of $H_{dot}$
describes the hopping coupling between the two dots with $J$ being the
hopping parameter. The two metallic electrodes are modeled as
non-interacting Fermi gases and the corresponding Hamiltonians are given by
\begin{equation}
H_{\text{Leads,2}}=\sum_{\alpha \mathbf{k}}\varepsilon _{\alpha \mathbf{k}%
}a_{\alpha \mathbf{k}}^{\dag }a_{\alpha \mathbf{k}},  \label{Leads2}
\end{equation}%
where $a_{\alpha \mathbf{k}}^{\dag }$ ($a_{\alpha \mathbf{k}}$) creates
(annihilates) an electron with energy $\varepsilon _{\alpha \mathbf{k}}$ and
momentum $\mathbf{k}$ in $\alpha $ ($\alpha =L,R$) electrode. The tunneling
between the double QDs and the two electrodes is described by
\begin{equation}
H_{\text{T,2}}=\sum_{\alpha \mathbf{k}}\left( t_{L}a_{L\mathbf{k}}^{\dag
}d_{1}+t_{R}a_{R\mathbf{k}}^{\dag }d_{2}+\text{H.c.}\right) .
\label{tunneling2}
\end{equation}%
For the case of the weak QD-electrode coupling, the particle-number-resolved
TCL quantum master equation for the reduced density matrix of the considered
serially double-QD system reads
\begin{eqnarray}
&&\left. \dot{\rho}^{\left( n\right) }\left( t\right) \right\vert _{\text{%
dot,2}}  \notag \\
&=&-i\mathcal{L}\rho ^{\left( n\right) }\left( t\right) -\left[
d_{1}^{\dagger }A_{L}^{\left( -\right) }\rho ^{\left( n\right) }\left(
t\right) +\rho ^{\left( n\right) }\left( t\right) A_{L}^{\left( +\right)
}d_{1}^{\dagger }\right.   \notag \\
&&d_{2}^{\dagger }A_{R}^{\left( -\right) }\rho ^{\left( n\right) }\left(
t\right) +\rho ^{\left( n\right) }\left( t\right) A_{R}^{\left( +\right)
}d_{2}^{\dagger }-A_{L}^{\left( -\right) }\rho ^{\left( n\right) }\left(
t\right) d_{1}^{\dagger }  \notag \\
&&-d_{1}^{\dagger }\rho ^{\left( n\right) }\left( t\right) A_{L}^{\left(
+\right) }-A_{R}^{\left( -\right) }\rho ^{\left( n-1\right) }\left( t\right)
d_{2}^{\dagger }  \notag \\
&&\left. -d_{2}^{\dagger }\rho ^{\left( n+1\right) }\left( t\right)
A_{R}^{\left( +\right) }+\text{H.c.}\right] ,  \label{Master2}
\end{eqnarray}%
Here, we can diagonalize the serially coupled\ double QDs Hamiltonian $H_{%
\text{dot,2}}$ in the basis represented by the electron occupation numbers
in the QD-1 and QD-2 denoted respectively by $N_{L}$ and $N_{R}$, namely, $%
\left\{ \left\vert 0,0\right\rangle ,\left\vert 1,0\right\rangle ,\left\vert
0,1\right\rangle ,\left\vert 1,1\right\rangle \right\} $, and obtain the
corresponding four eigenstates of the considered serially coupled double QDs
system\cite{Xueepjb}
\begin{align}
H_{\text{dot,2}}\left\vert 0\right\rangle & =0,\left\vert 0\right\rangle
=\left\vert 0,0\right\rangle ,  \notag \\
H_{\text{dot,2}}\left\vert 1\right\rangle ^{\pm }& =\epsilon _{\pm
}\left\vert 1\right\rangle ^{\pm },\left\vert 1\right\rangle ^{\pm }=a_{\pm
}\left\vert 1,0\right\rangle +b_{\pm }\left\vert 0,1\right\rangle ,
\label{eigenstates} \\
H_{\text{dot,2}}\left\vert 2\right\rangle & =\epsilon _{2}\left\vert
2\right\rangle =\left( \epsilon _{1}+\epsilon _{2}+U\right) \left\vert
2\right\rangle ,\left\vert 2\right\rangle =\left\vert 1,1\right\rangle ,
\notag
\end{align}%
with
\begin{equation}
\epsilon _{\pm }=\frac{\left( \epsilon _{1}+\epsilon _{1}\right) \pm \sqrt{%
\left( \epsilon _{1}-\epsilon _{1}\right) ^{2}+4J^{2}}}{2},
\label{eigenvalues}
\end{equation}%
and%
\begin{eqnarray}
a_{\pm } &=&\frac{\mp J}{\sqrt{\left( \epsilon _{\pm }-\epsilon _{1}\right)
^{2}+J^{2}}},  \notag \\
b_{\pm } &=&\frac{\pm \left( \epsilon _{\pm }-\epsilon _{1}\right) }{\sqrt{%
\left( \epsilon _{\pm }-\epsilon _{1}\right) ^{2}+J^{2}}}.  \label{abaddsub}
\end{eqnarray}%
Here, we focus on the regime $\left( \epsilon _{+}-\epsilon _{-}\right) \ll
k_{B}T$, where the hopping coupling between the two QDs strongly modifies
the internal dynamics, and the off-diagonal elements of the reduced density
matrix play an essential role in the electron tunneling processes\cite%
{Gurvitz,Stoof96,Aguado04,LuoJY11}. In the following numerical
calculations, thus, the parameters of the serially coupled double QDs system
are chosen as $\epsilon _{1}=\epsilon _{2}=1$, $J=0.001$, $U=4$ and $%
k_{B}T=0.05$.

When the coupling of the QD-2 with the right (drain) electrode is stronger than
that of the QD-1 with the left (source) electrode, namely, $\Gamma _{L}/\Gamma
_{R}<1$, we plot the first four current cumulants as a function of the bias
voltage for different values of the QD-2-electrode coupling $\Gamma _{R}$ at
$\Gamma _{L}/\Gamma _{R}=0.1$ in Figs. 3(a)-3(d). We found that the
non-Markovian effect has a very weak influence on the FCS. Interestingly,
the high-order current cumulants the skewness and the kurtosis can still
show the tiny differences, see Figs. 3(c) and 3(d). Whereas for
the $\Gamma _{L}/\Gamma _{R}\geq 1$ case, the
non-Markovian effect has a significant impact on the FCS, see Fig. 4.
Especially, for a relatively large value of the ratio $\Gamma _{L}/\Gamma
_{R}=10$ and the coupling of the QD-1 with the left electrode being
stronger than the hoping coupling, namely, $\Gamma _{L}/J>1$, the
non-Markovian effect can induce a strong negative differential conductance
(NDC) and super-Poissonian noise, see Figs. 4(e) and 4(f). In addition, in
the case of $\Gamma _{L}/\Gamma _{R}\geq 1$ and $\Gamma _{L}/J>1$, the
transitions of the skewness and the kurtosis from positive (negative) to
negative (positive) values are observed, see the dotted line in Fig. 4(c),
the dotted and dash-dot-dotted lines in Fig. 4(d), and the dash-dot-dotted
line in Fig. 4(h). It is well known that the skewness and the kurtosis (both
its magnitude and sign) characterize, respectively, the asymmetry of and the
peakedness of the distribution around the average transferred-electron
number $\bar{n}$ during a time interval $t$, thus that provides further
information for the counting statistics beyond the shot noise.

To discuss the underlying mechanisms of the current noise clearly, for the
system parameters considered here, the two
singly-occupied eigenstates and eigenvalues can be expressed as
\begin{equation}
\left\{
\begin{array}{c}
\left\vert 1\right\rangle ^{\pm }=\mp \frac{\sqrt{2}}{2}\left\vert
1,0\right\rangle +\frac{\sqrt{2}}{2}\left\vert 0,1\right\rangle , \\
\epsilon _{+}=\epsilon _{-}=\epsilon .%
\end{array}%
\right.  \label{eigenmodified}
\end{equation}%
Here we have utilized the equations $\epsilon _{1}=\epsilon _{2}=\epsilon $
and $\epsilon \gg J$. In this situation, the equations of motion of the six
elements of the reduced density matrix are given by
\begin{align}
&  \left\langle 0,0\right\vert \dot{\rho}_{S}\left(  t\right)  \left\vert
0,0\right\rangle \nonumber\\
&  =-\left[  \Gamma_{L}f_{L,+}\left(  \epsilon\right)  +\Gamma_{R}%
f_{R,+}\left(  \epsilon\right)  \right]  \left\langle 0,0\right\vert \rho
_{S}\left(  t\right)  \left\vert 0,0\right\rangle \nonumber\\
&  +\frac{1}{2}\left[  \Gamma_{L}f_{L,-}\left(  \epsilon\right)  +\Gamma
_{R}f_{R,-}\left(  \epsilon\right)  e^{i\chi}\right]  \left\langle
1\right\vert ^{+}\rho_{S}\left(  t\right)  \left\vert 1\right\rangle
^{+}\nonumber\\
&  -\frac{1}{2}\left[  \Gamma_{L}f_{L,-}\left(  \epsilon\right)  -\Gamma
_{R}f_{R,-}\left(  \epsilon\right)  e^{i\chi}\right]  \left\langle
1\right\vert ^{+}\rho_{S}\left(  t\right)  \left\vert 1\right\rangle
^{-}\nonumber\\
&  -\frac{1}{2}\left[  \Gamma_{L}f_{L,-}\left(  \epsilon\right)  -\Gamma
_{R}f_{R,-}\left(  \epsilon\right)  e^{i\chi}\right]  \left\langle
1\right\vert ^{-}\rho_{S}\left(  t\right)  \left\vert 1\right\rangle
^{+}\nonumber\\
&  +\frac{1}{2}\left[  \Gamma_{L}f_{L,-}\left(  \epsilon\right)  +\Gamma
_{R}f_{R,-}\left(  \epsilon\right)  e^{i\chi}\right]  \left\langle
1\right\vert ^{-}\rho_{S}\left(  t\right)  \left\vert 1\right\rangle
^{-} , \label{zero2}%
\end{align}%
\begin{align}
&  \left\langle 1\right\vert ^{\pm}\dot{\rho}_{S}^{\left(  n\right)  }\left(
t\right)  \left\vert 1\right\rangle ^{\pm}\nonumber\\
&  =\frac{1}{2}\left[  \Gamma_{L}f_{L,+}\left(  \epsilon\right)  +\Gamma
_{R}f_{R,+}\left(  \epsilon\right)  e^{-i\chi}\right]  \left\langle
0,0\right\vert \rho_{S}\left(  t\right)  \left\vert 0,0\right\rangle
\nonumber\\
&  -\frac{1}{2}%
{\displaystyle\sum\limits_{\alpha=L,R}}
\Gamma_{\alpha}\left[  f_{\alpha,+}\left(  \epsilon+U\right)  +f_{\alpha
,-}\left(  \epsilon\right)  \right]  \left\langle 1\right\vert ^{\pm}\rho
_{S}\left(  t\right)  \left\vert 1\right\rangle ^{\pm}\nonumber\\
&  \pm\frac{1}{2}\frac{\Gamma_{L}}{2\pi}\left(  i\Phi_{L}\mp\pi F_{L}\right)
\left\langle 1\right\vert ^{+}\rho_{S}\left(  t\right)  \left\vert
1\right\rangle ^{-}\mp\frac{1}{2}\frac{\Gamma_{R}}{2\pi}\left(  i\Phi_{R}%
\mp\pi F_{R}\right)  \left\langle 1\right\vert ^{+}\rho_{S}\left(  t\right)
\left\vert 1\right\rangle ^{-}\nonumber\\
&  \mp\frac{1}{2}\frac{\Gamma_{\alpha}}{2\pi}\left(  i\Phi_{L}\pm\pi
F_{L}\right)  \left\langle 1\right\vert ^{-}\rho_{S}\left(  t\right)
\left\vert 1\right\rangle ^{+}\pm\frac{1}{2}\frac{\Gamma_{R}}{2\pi}\left(
i\Phi_{R}\pm\pi F_{R}\right)  \left\langle 1\right\vert ^{-}\rho_{S}\left(
t\right)  \left\vert 1\right\rangle ^{+}\nonumber\\
&  +\frac{1}{2}\left[  \Gamma_{L}f_{L,-}\left(  \epsilon+U\right)  +\Gamma
_{R}f_{R,-}\left(  \epsilon+U\right)  e^{i\chi}\right]  \left\langle
1,1\right\vert \rho_{S}\left(  t\right)  \left\vert 1,1\right\rangle ,
\label{add2}%
\end{align}%
\begin{align}
&  \left\langle 1\right\vert ^{\pm}\dot{\rho}_{S}\left(  t\right)  \left\vert
1\right\rangle ^{\mp}\nonumber\\
&  =-\frac{1}{2}\left[  \Gamma_{L}f_{L,+}\left(  \epsilon\right)  -\Gamma
_{R}f_{R,+}\left(  \epsilon\right)  e^{-i\chi}\right]  \left\langle
0,0\right\vert \rho_{S}\left(  t\right)  \left\vert 0,0\right\rangle
\nonumber\\
&  \pm\frac{1}{2}\frac{\Gamma_{L}}{2\pi}\left(  i\Phi_{L}\mp\pi F_{L}\right)
\left\langle 1\right\vert ^{+}\rho_{S}^{\left(  n\right)  }\left(  t\right)
\left\vert 1\right\rangle ^{+}\mp\frac{1}{2}\frac{\Gamma_{R}}{2\pi}\left(
i\Phi_{R}\mp\pi F_{R}\right)  \left\langle 1\right\vert ^{+}\rho_{S}\left(
t\right)  \left\vert 1\right\rangle ^{+}\nonumber\\
&  -\frac{1}{2}%
{\displaystyle\sum\limits_{\alpha=L,R}}
\Gamma_{\alpha}\left[  f_{\alpha,+}\left(  \epsilon+U\right)  +f_{\alpha
,-}\left(  \epsilon\right)  \right]  \left\langle 1\right\vert ^{\pm}\rho
_{S}\left(  t\right)  \left\vert 1\right\rangle ^{\mp}\mp2iJ\left\langle
1\right\vert ^{\pm}\rho_{S}\left(  t\right)  \left\vert 1\right\rangle ^{\mp
}\nonumber\\
&  \mp\frac{1}{2}\frac{\Gamma_{L}}{2\pi}\left(  i\Phi_{L}\pm\pi F_{L}\right)
\left\langle 1\right\vert ^{-}\rho_{S}^{\left(  n\right)  }\left(  t\right)
\left\vert 1\right\rangle ^{-}\pm\frac{1}{2}\frac{\Gamma_{R}}{2\pi}\left(
i\Phi_{R}\pm\pi F_{R}\right)  \left\langle 1\right\vert ^{-}\rho_{S}\left(
t\right)  \left\vert 1\right\rangle ^{-}\nonumber\\
&  +\frac{1}{2}\left[  \Gamma_{L}f_{L,-}\left(  \epsilon+U\right)  -\Gamma
_{R}f_{R,-}\left(  \epsilon+U\right)  e^{i\chi}\right]  \left\langle
1,1\right\vert \rho_{S}\left(  t\right)  \left\vert 1,1\right\rangle ,
\label{addsub2}%
\end{align}%
\begin{align}
&  \left\langle 1,1\right\vert \dot{\rho}_{S}\left(  t\right)  \left\vert
1,1\right\rangle \nonumber\\
&  =\frac{1}{2}\left[  \Gamma_{L}f_{L,+}\left(  \epsilon+U\right)  +\Gamma
_{R}f_{R,+}\left(  \epsilon+U\right)  e^{-i\chi}\right]  \left\langle
1\right\vert ^{+}\rho_{S}\left(  t\right)  \left\vert 1\right\rangle
^{+}\nonumber\\
&  +\frac{1}{2}\left[  \Gamma_{L}f_{L,+}\left(  \epsilon+U\right)  -\Gamma
_{R}f_{R,+}\left(  \epsilon+U\right)  e^{-i\chi}\right]  \left\langle
1\right\vert ^{+}\rho_{S}\left(  t\right)  \left\vert 1\right\rangle
^{-}\nonumber\\
&  +\frac{1}{2}\left[  \Gamma_{L}f_{L,+}\left(  \epsilon+U\right)  -\Gamma
_{R}f_{R,+}\left(  \epsilon+U\right)  e^{-i\chi}\right]  \left\langle
1\right\vert ^{-}\rho_{S}\left(  t\right)  \left\vert 1\right\rangle
^{+}\nonumber\\
&  +\frac{1}{2}\left[  \Gamma_{L}f_{L,+}\left(  \epsilon+U\right)  +\Gamma
_{R}f_{R,+}\left(  \epsilon+U\right)  e^{-i\chi}\right]  \left\langle
1\right\vert ^{-}\rho_{S}\left(  t\right)  \left\vert 1\right\rangle
^{-}\nonumber\\
&  -\left[  \Gamma_{L}f_{L,-}\left(  \epsilon+U\right)  +\Gamma_{R}%
f_{R,-}\left(  \epsilon+U\right)  \right]  \left\langle 1,1\right\vert
\rho_{S}\left(  t\right)  \left\vert 1,1\right\rangle ,  \label{two2}%
\end{align}
where $\Phi _{\alpha }=\phi _{\alpha }\left( \epsilon +U\right) -\phi
_{\alpha }\left( \epsilon \right) $, $\phi _{\alpha }\left( \epsilon \right)
=\operatorname{Re}\Psi \left[ \frac{1}{2}+\frac{i\left( \epsilon -\mu _{L}\right) }{%
2\pi k_{B}T}\right] $ ($\Psi $ is the digamma function) and $F_{\alpha }=f_{\alpha ,+}\left( \epsilon
+U\right) -f_{\alpha ,-}\left( \epsilon \right) $. Compared with the
Markovian case, it is obvious that the non-Markovian effect manifests itself
through the off-diagonal elements of the reduced density matrix, namely, the
quantum coherence of the considered QDs system. In Fig. 5(a), we plot the
functions $\Phi _{L}-0.1\Phi _{R}$ $\left( \Gamma _{R}=0.1\Gamma _{L}\right)
$, $\Phi _{L}-\Phi _{R}$ $\left( \Gamma _{R}=\Gamma _{L}\right) $ and $%
0.1\Phi _{L}-\Phi _{R}$ $\left( \Gamma _{L}=0.1\Gamma _{R}\right) $ as a
function of bias voltage. It is clearly evident that the values of the
functions $\Phi _{L}-0.1\Phi _{R}$ and $\Phi _{L}-\Phi _{R}$ show
significant variations with increasing bias voltage, especially in the
vicinity of the bias voltages $V_{b}=2$ and $V_{b}=10$ because the new transport channels begin to
participate in quantum transport; while $0.1\Phi _{L}-\Phi _{R}$
has a gentle variation. Consequently, the non-Markovian effects in
the $\Gamma _{L}/\Gamma _{R}\geq 1$ case have a remarkable impact on the
FCS, see Fig. 4. Moreover, for $\Gamma _{L}/\Gamma _{R}=10$ case, the non-Markovian effect has a more
significant on the FCS than the $\Gamma _{L}/\Gamma _{R}=1$ case, which
originates from the QD-2-electrode coupling $\Gamma _{R}$ is weaker than the
hoping coupling $J$, where the electron tunneling from QD-1 can not tunnel
out QD-2 very quickly and still influence the internal dynamics.

In order to illustrate whether the non-Markovian effect has a weak influence
on the FCS in a relatively small quantum-coherent QD system, we
consider the regime $\left( \epsilon _{+}-\epsilon _{-}\right) \gg k_{B}T$ $%
(J=1)$, where the off-diagonal elements of the reduced density matrix have
little influence on the electron tunneling processes. We find that for the $%
J=1$ case the diagonal elements of the reduced density matrix play a major
role in the electron tunneling processes, and the non-Markovian effect in
this case indeed has little impact on the FCS, see Figs. 3(e)-3(h). Consequently,
the influence of the non-Markovian effect on the FCS depends on the quantum
coherence of the considered QD system. To prove whether this conclusion
is universal or not, we take side-coupled double QDs for further
illustration in the following subsection.

\subsection*{Side-coupled double quantum dots with high quantum coherence}

We consider here a side-coupled double QDs system. In this case, the QD-1 is
only weakly coupled to the two electrodes, see Fig. 1(b). The QD-electrode
tunneling is thus described by
\begin{equation}
H_{\text{T,3}}=\sum_{\alpha \mathbf{k}}\left( t_{\alpha \mathbf{k}}a_{\alpha
\mathbf{k}}^{\dag }d_{1}+t_{\alpha \mathbf{k}}^{\ast }d_{1}^{\dag }a_{\alpha
\mathbf{k}}\right) .  \label{tunneling3}
\end{equation}

In the case of the QD-electrode weak coupling, the particle-number-resolved
TCL quantum master equation for the side-coupled double QDs can be expressed
as
\begin{eqnarray}
&&\left. \dot{\rho}^{\left( n\right) }\left( t\right) \right\vert _{\text{%
dot,3}}  \notag \\
&=&-i\mathcal{L}\rho ^{\left( n\right) }\left( t\right) -\left[
d_{1}^{\dagger }A_{L,1}^{\left( -\right) }\rho ^{\left( n\right) }\left(
t\right) +d_{1}^{\dagger }A_{R,1}^{\left( -\right) }\rho ^{\left( n\right)
}\left( t\right) \right.   \notag \\
&&+\rho ^{\left( n\right) }\left( t\right) A_{L,1}^{\left( +\right)
}d_{1}^{\dagger }+\rho ^{\left( n\right) }\left( t\right) A_{R,1}^{\left(
+\right) }d_{1}^{\dagger }-A_{L,1}^{\left( -\right) }\rho ^{\left( n\right)
}\left( t\right) d_{1}^{\dagger }  \notag \\
&&-A_{R,1}^{\left( -\right) }\rho ^{\left( n-1\right) }\left( t\right)
d_{1}^{\dagger }-d_{1}^{\dagger }\rho ^{\left( n\right) }\left( t\right)
A_{L,1}^{\left( +\right) }  \notag \\
&&\left. -d_{1}^{\dagger }\rho ^{\left( n+1\right) }\left( t\right)
A_{R,1}^{\left( +\right) }+\text{H.c.}\right] .  \label{Master3}
\end{eqnarray}%
Here, the eigenstates and eigenvalues of the side-coupled double QDs system
are the same as the serially coupled double QDs system. In the following
numerical calculations, the parameters of the side-coupled QDs system are
chosen as $\epsilon _{1}=\epsilon _{2}=1$, $J=0.001$, $U=5$ and $k_{B}T=0.1$.

For the present side-coupled QDs system with high quantum coherence, we
find that for $\Gamma _{L}/\Gamma _{R}\geq 1$ case the non-Markovian effect
has a more remarkable impact on the FCS than that in the serially coupled
double QDs system, but the NDC does not appear, see Figs. 4 and 6. For
instance, in the case of $\Gamma _{L}/J>1$ and $\Gamma _{L}/\Gamma _{R}=1$,
the non-Markovian effect can further enhance the super-Poissonian
shot noise, see the dotted and dash-dot-dotted lines in Fig. 6(b); and the
transitions of the skewness and the kurtosis from a relatively small
positive to a large negative values take place, especially for a relatively
large value $\Gamma _{L}/J$ the kurtosis can be further decreased to a very
large negative value, see the dotted and dash-dot-dotted lines in Figs. 6(c)
and 6(d). While for the $\Gamma _{L}/J>1$ and $\Gamma _{L}/\Gamma _{R}=10$
case the non-Markovian effect can enhance the shot noise to
a super-Poissonian value, see the dotted and dash-dot-dotted lines in
Fig. 6(f), and the transition of the kurtosis from small positive to large
negative values only takes place, see the dotted and dash-dot-dotted lines
in Fig. 6(h). For the system parameters considered here, namely, in the
limit of $\epsilon \gg J$, the equations of motion of the six
elements of the reduced density matrix read
\begin{align}
&  \left\langle 0,0\right\vert \dot{\rho}_{S}\left(  t\right)  \left\vert
0,0\right\rangle \nonumber\\
&  =-\left[  \Gamma_{L}f_{L,+}\left(  \epsilon\right)  +\Gamma_{R}%
f_{R,+}\left(  \epsilon\right)  \right]  \left\langle 0,0\right\vert \rho
_{S}\left(  t\right)  \left\vert 0,0\right\rangle \nonumber\\
&  +\frac{1}{2}\left[  \Gamma_{L}f_{L,-}\left(  \epsilon\right)  +\Gamma
_{R}f_{R,-}\left(  \epsilon\right)  e^{i\chi}\right]  \left\langle
1\right\vert ^{+}\rho_{S}\left(  t\right)  \left\vert 1\right\rangle
^{+}\nonumber\\
&  -\frac{1}{2}\left[  \Gamma_{L}f_{L,-}\left(  \epsilon\right)  +\Gamma
_{R}f_{R,-}\left(  \epsilon\right)  e^{i\chi}\right]  \left\langle
1\right\vert ^{+}\rho_{S}\left(  t\right)  \left\vert 1\right\rangle
^{-}\nonumber\\
&  -\frac{1}{2}\left[  \Gamma_{L}f_{L,-}\left(  \epsilon\right)  +\Gamma
_{R}f_{R,-}\left(  \epsilon\right)  e^{i\chi}\right]  \left\langle
1\right\vert ^{-}\rho_{S}\left(  t\right)  \left\vert 1\right\rangle
^{+}\nonumber\\
&  +\frac{1}{2}\left[  \Gamma_{L}f_{L,-}\left(  \epsilon\right)  +\Gamma
_{R}f_{R,-}\left(  \epsilon\right)  e^{i\chi}\right]  \left\langle
1\right\vert ^{-}\rho_{S}\left(  t\right)  \left\vert 1\right\rangle
^{-} , \label{zero3}%
\end{align}%
\begin{align}
&  \left\langle 1\right\vert ^{\pm}\dot{\rho}_{S}\left(  t\right)  \left\vert
1\right\rangle ^{\pm}\nonumber\\
&  =\frac{1}{2}\left[  \Gamma_{L}f_{L,+}\left(  \epsilon\right)  +\Gamma
_{R}f_{R,+}\left(  \epsilon\right)  e^{-i\chi}\right]  \left\langle
0,0\right\vert \rho_{S}\left(  t\right)  \left\vert 0,0\right\rangle
\nonumber\\
&  -\frac{1}{2}%
{\displaystyle\sum\limits_{\alpha=L,R}}
\Gamma_{\alpha}\left[  f_{\alpha,+}\left(  \epsilon+U\right)  +f_{\alpha
,-}\left(  \epsilon\right)  \right]  \left\langle 1\right\vert ^{\pm}\rho
_{S}\left(  t\right)  \left\vert 1\right\rangle ^{\pm}\nonumber\\
&  \pm\frac{1}{2}%
{\displaystyle\sum\limits_{\alpha=L,R}}
\frac{\Gamma_{\alpha}}{2\pi}\left(  i\Phi_{\alpha}\mp\pi F_{\alpha}\right)
\left\langle 1\right\vert ^{+}\rho_{S}\left(  t\right)  \left\vert
1\right\rangle ^{-}\nonumber\\
&  \mp\frac{1}{2}%
{\displaystyle\sum\limits_{\alpha=L,R}}
\frac{\Gamma_{\alpha}}{2\pi}\left[  i\Phi_{\alpha}\pm\pi F_{\alpha}\right]
\left\langle 1\right\vert ^{-}\rho_{S}\left(  t\right)  \left\vert
1\right\rangle ^{+}\nonumber\\
&  +\frac{1}{2}\left[  \Gamma_{L}f_{L,-}\left(  \epsilon+U\right)  +\frac
{1}{2}\Gamma_{R}f_{R,-}\left(  \epsilon+U\right)  e^{i\chi}\right]
\left\langle 1,1\right\vert \rho_{S}\left(  t\right)  \left\vert
1,1\right\rangle , \label{add3}%
\end{align}%
\begin{align}
&  \left\langle 1\right\vert ^{\pm}\dot{\rho}_{S}\left(  t\right)  \left\vert
1\right\rangle ^{\mp}\nonumber\\
&  =-\frac{1}{2}\left[  \Gamma_{L}f_{L,+}\left(  \epsilon\right)  +\Gamma
_{R}f_{R,+}\left(  \epsilon\right)  e^{-i\chi}\right]  \left\langle
0,0\right\vert \rho_{S}\left(  t\right)  \left\vert 0,0\right\rangle
\nonumber\\
&  \pm\frac{1}{2}%
{\displaystyle\sum\limits_{\alpha=L,R}}
\frac{\Gamma_{\alpha}}{2\pi}\left[  i\Phi_{\alpha}\mp\pi F_{\alpha}\right]
\left\langle 1\right\vert ^{+}\rho_{S}^{\left(  n\right)  }\left(  t\right)
\left\vert 1\right\rangle ^{+}\nonumber\\
&  -\frac{1}{2}%
{\displaystyle\sum\limits_{\alpha=L,R}}
\Gamma_{\alpha}\left[  f_{\alpha,+}\left(  \epsilon+U\right)  +f_{\alpha
,-}\left(  \epsilon\right)  \right]  \left\langle 1\right\vert ^{\pm}\rho
_{S}\left(  t\right)  \left\vert 1\right\rangle ^{\mp}\mp2iJ\left\langle
1\right\vert ^{\pm}\rho_{S}\left(  t\right)  \left\vert 1\right\rangle ^{\mp
}\nonumber\\
&  \mp\frac{1}{2}%
{\displaystyle\sum\limits_{\alpha=L,R}}
\frac{\Gamma_{\alpha}}{2\pi}\left[  i\Phi_{\alpha}\pm\pi F_{\alpha}\right]
\left\langle 1\right\vert ^{-}\rho_{S}^{\left(  n\right)  }\left(  t\right)
\left\vert 1\right\rangle ^{-}\nonumber\\
&  +\frac{1}{2}\left[  \Gamma_{L}f_{L,-}\left(  \epsilon+U\right)  +\Gamma
_{R}f_{R,-}\left(  \epsilon+U\right)  e^{i\chi}\right]  \left\langle
1,1\right\vert \rho_{S}\left(  t\right)  \left\vert 1,1\right\rangle ,
\label{addsub3}%
\end{align}%
\begin{align}
&  \left\langle 1,1\right\vert \dot{\rho}_{S}\left(  t\right)  \left\vert
1,1\right\rangle \nonumber\\
&  =\frac{1}{2}\left[  \Gamma_{L}f_{L,+}\left(  \epsilon+U\right)  +\Gamma
_{R}f_{R,+}\left(  \epsilon+U\right)  e^{-i\chi}\right]  \left\langle
1\right\vert ^{+}\rho_{S}\left(  t\right)  \left\vert 1\right\rangle
^{+}\nonumber\\
&  +\frac{1}{2}\left[  \Gamma_{L}f_{L,+}\left(  \epsilon+U\right)  +\Gamma
_{R}f_{R,+}\left(  \epsilon+U\right)  e^{-i\chi}\right]  \left\langle
1\right\vert ^{+}\rho_{S}\left(  t\right)  \left\vert 1\right\rangle
^{-}\nonumber\\
&  +\frac{1}{2}\left[  \Gamma_{L}f_{L,+}\left(  \epsilon+U\right)  +\Gamma
_{R}f_{R,+}\left(  \epsilon+U\right)  e^{-i\chi}\right]  \left\langle
1\right\vert ^{-}\rho_{S}\left(  t\right)  \left\vert 1\right\rangle
^{+}\nonumber\\
&  +\frac{1}{2}\left[  \Gamma_{L}f_{L,+}\left(  \epsilon+U\right)  +\Gamma
_{R}f_{R,+}\left(  \epsilon+U\right)  e^{-i\chi}\right]  \left\langle
1\right\vert ^{-}\rho_{S}\left(  t\right)  \left\vert 1\right\rangle
^{-}\nonumber\\
&  -\left[  \Gamma_{L}f_{L,-}\left(  \epsilon+U\right)  +\Gamma_{R}%
f_{R,-}\left(  \epsilon+U\right)  \right]  \left\langle 1,1\right\vert
\rho_{S}\left(  t\right)  \left\vert 1,1\right\rangle . \label{two3}%
\end{align}
From the above four equations, we find that these characteristics also originate
from the quantum coherence of the side-coupled double QDs, and can also be
understood in terms of the functions $\Phi _{L}+0.1\Phi _{R}$ and $%
\Phi _{L}+\Phi _{R}$, which have considerable variations in the vicinity of the
bias voltages $V_{b}=2$ and $V_{b}=12$ because the new transport channels begin to enter the bias
voltage window, see the solid and dashed lines in Fig. 5(b). As for the $%
\Gamma _{L}/\Gamma _{R}<1$ case the non-Markovian effect has a slightly
influence on the FCS because the function $0.1\Phi _{L}+\Phi _{R}$ has a
gentle variation with increasing the bias voltage, see the dotted line in
Fig. 5(b), which is the same as the serially coupled double QDs system, see
Figs. 3(a)-3(d) and 7.

In addition, it should be pointed out that for $\Gamma _{L}/\Gamma _{R}=1$
the non-Markovian effect has a stronger impact on the FCS than that for $%
\Gamma _{L}/\Gamma _{R}>1$ case, which is contrary to the case of the
serially coupled double QDs system. For the the side-coupled double QDs
system, the quantum coherence originates from the quantum interference
between the direct electron tunneling process, namely, the
conduction-electron tunneling into the QD-1 and then directly tunneling out of the
QD-1 onto the drain electrode, and the indirect tunneling process, namely, the
conduction-electron from the source electrode first tunneling from the QD-1 to
the QD-2, then tunneling back into the QD-1, and at last tunneling out of the
QD-1. Thus, the fast direct tunneling process in the $\Gamma _{L}=10\Gamma _{R}$
case can be suppressed compared with the $\Gamma _{L}=\Gamma _{R}$ case,
which leads to the non-Markovian effect has a relatively strong impact on
the FCS in the $\Gamma _{L}/\Gamma _{R}=1$ case.

\section*{Discussion}
We have developed a non-Markovian FCS formalism based on the exact TCL
master equation, and studied the influence of the interplay between the
quantum coherence and non-Markovian effect on the long-time limit of the FCS
in three QD systems, namely, single QD, serially coupled double QDs
and side-coupled double QDs. It is demonstrated that the
non-Markovian effect manifests itself through the quantum coherence of the
considered QD molecule system, and especially has a significant impact on
the FCS in the high quantum-coherent QD molecule system, which depends on the
coupling of the considered QD molecule system with the source and drain electrodes.
For the single QD system without quantum coherence, the non-Markovian effect
has no influence on the current noise properties; whereas for the serially
coupled and side-coupled double QDs systems with high quantum coherence,
that has a remarkable impact on the FCS when the coupling of the considered
QD molecule with the incident electrode is equal to or stronger
than that with the outgoing electrode. For instance, for the high quantum-coherent serially
coupled double QDs system, the non-Markovian effect can induce a strong NDC and
change the shot noise from the sub-Poissonian to super-Poissonian distribution in the case of $%
\Gamma _{L}/\Gamma _{R}\gg 1$ and $\Gamma _{L}>J$; while for the high quantum-coherent side-coupled
double QDs system, that can remarkably enhance the super-Poissonian noise or the sub-Poissonian noise
for the $\Gamma _{L}/\Gamma _{R}\geq 1$ case. Moreover, the non-Markovian
effect can also lead to the occurrences of the skewness and kurtosis from
small positive to large negative values. These results indicated
that the influence of the non-Markovian effect on the long-time limit of the
FCS should be considered in a highly quantum-coherent single-molecule system.

\section*{Methods}

\subsection*{Particle-number-resolved time-convolutionless quantum master equation}
We consider a general transport setup consisting of a single-level QD molecule
weakly coupled to the two electrodes, see Fig. 1, which is described by the
following Hamiltonian
\begin{equation}
H=H_{\text{electrodes}}+H_{\text{dot}}+H_{\text{hyb}}.  \label{Hamiltonian}
\end{equation}%
Here, the first term $H_{\text{electrodes}}=\sum_{\alpha ,k,\sigma
}\varepsilon _{\alpha k}a_{\alpha k\sigma }^{\dag }a_{\alpha k\sigma }$
stands for the Hamiltonians of the two electrodes, with $\varepsilon
_{\alpha k}$ being the energy dispersion, and $a_{\alpha k\sigma }$ $%
(a_{\alpha k\sigma }^{\dag })$ the annihilation (creation) operators in the $%
\alpha $ electrode. The second term $H_{\text{dot}}=H_{S}\left( d_{\mu
}^{\dag },d_{\mu }\right) $, which may contain vibrational or spin degrees
of freedom and different types of many-body interaction, represents the QD
molecule Hamiltonian, where $d_{\mu }^{\dag }$ $\left( d_{\mu }\right) $ is
the creation (annihilation) operator of electrons in a quantum state denoted
by $\mu $. The third term $H_{\text{hyb}}=\sum_{\alpha ,\mu ,k}\left(
t_{\alpha \mu k}^{\ast }a_{\alpha \mu k}^{\dag }d_{\mu }+t_{\alpha \mu
k}d_{\mu }^{\dag }a_{\alpha \mu k}\right) $ describes the tunneling coupling
between the QD molecule and the two electrodes, which is assumed to be a
sum of bilinear terms that each create an electron in the QD molecule and
annihilate one in the electrodes or vice versa.

The QD-electrode coupling is assumed to be sufficiently weak, so that $H_{%
\text{hyb}}$ can be treated perturbatively. In the interaction
representation, the equation of motion for the total density matrix reads
\begin{equation}
\frac{\partial }{\partial t}\rho ^{I}\left( t\right) =-i\left[ H_{\text{hyb}%
}^{I}\left( t\right) ,\rho ^{I}\left( t\right) \right] \equiv \mathcal{L}%
\left( t\right) \rho ^{I}\left( t\right) ,  \label{interaction}
\end{equation}%
with
\begin{equation*}
H_{\text{hyb}}^{I}\left( t\right) =-\sum_{\alpha ,\mu }\left[ f_{\alpha \mu
}^{\dag }\left( t\right) d_{\mu }\left( t\right) +d_{\mu }^{\dag }\left(
t\right) f_{\alpha \mu }\left( t\right) \right]
\end{equation*}%
where $f_{\alpha \mu }^{\dag }\left( t\right) =\sum_{k}t_{\alpha \mu
k}^{\ast }e^{iH_{\text{electrodes}}t}a_{\alpha \mu k}^{\dag }e^{-iH_{\text{%
electrodes}}t}$ and $d_{\mu }\left( t\right) =e^{iH_{\text{dot}}t}d_{\mu
}e^{-iH_{\text{dot}}t}$. In order to derive an exact equation of motion for
the reduced density matrix $\rho _{S}$ of the QD molecule system, it is
convenient to define a super-operator $\mathcal{P}$ according to
\begin{equation}
\mathcal{P}\rho =\text{tr}_{B}\left[ \rho \right] \otimes \rho _{B}=\rho
_{S}\otimes \rho _{B},  \label{Pdefinition}
\end{equation}%
with $\rho _{B}$ being some fixed state of the electron electrode.
Accordingly, a complementary super-operator $\mathcal{Q}$ reads
\begin{equation}
\mathcal{Q}\rho =\rho -\mathcal{P}\rho .  \label{Qdefinition}
\end{equation}%
For a factorizing initial condition $\rho \left( t_{0}\right) =\rho
_{S}\left( t_{0}\right) \otimes \rho _{B}$, $\mathcal{P}\rho \left(
t_{0}\right) =\rho\left( t_{0}\right) $, and $\mathcal{Q}\rho \left(
t_{0}\right) =0$. Using the TCL projection operator method \cite{book}, one
can obtain the second-order TCL master equation
\begin{equation}
\frac{\partial }{\partial t}\mathcal{P}\rho \left( t\right) =\int_{-\infty
}^{t}dt_{1}\mathcal{PL}\left( t\right) \mathcal{L}\left( t_{1}\right)
\mathcal{P}\rho \left( t\right) ,  \label{second-order}
\end{equation}%
The Eq. (\ref{second-order}) is the starting point of deriving the
particle-number-resolved quantum master equation. Using Eqs. (\ref%
{interaction}) and (\ref{Pdefinition}), after some algebraic calculations
we can rewrite Eq. (\ref{second-order}) as
\begin{eqnarray}
&&\frac{\partial }{\partial t}\rho _{I,S}\left( t\right)  \notag \\
&=&-\sum_{\alpha \mu \nu }\int_{-\infty }^{t}dt_{1}\text{tr}_{B}\left[ \rho
_{I,S}\left( t\right) \otimes \rho _{B}f_{\alpha \nu }^{\dag }\left(
t_{1}\right) d_{\nu }\left( t_{1}\right) d_{\mu }^{\dag }\left( t\right)
f_{\alpha \mu }\left( t\right) \right]  \notag \\
&&-\sum_{\alpha \mu \nu }\int_{-\infty }^{t}dt_{1}\text{tr}_{B}\left[ d_{\mu
}^{\dag }\left( t\right) f_{\alpha \mu }\left( t\right) f_{\alpha \nu
}^{\dag }\left( t_{1}\right) d_{\nu }\left( t_{1}\right) \rho _{I,S}\left(
t\right) \otimes \rho _{B}\right]  \notag \\
&&+\sum_{\alpha \mu \nu }\int_{-\infty }^{t}dt_{1}\text{tr}_{B}\left[
f_{\alpha \mu }^{\dag }\left( t\right) d_{\mu }\left( t\right) \rho
_{I,S}\left( t\right) \otimes \rho _{B}d_{\nu }^{\dag }\left( t_{1}\right)
f_{\alpha \nu }\left( t_{1}\right) \right]  \notag \\
&&+\sum_{\alpha \mu \nu }\int_{-\infty }^{t}dt_{1}\text{tr}_{B}\left[ d_{\mu
}^{\dag }\left( t\right) f_{\alpha \mu }\left( t\right) \rho _{I,S}\left(
t\right) \otimes \rho _{B}f_{\alpha \nu }^{\dag }\left( t_{1}\right) d_{\nu
}\left( t_{1}\right) \right] +H.c..  \label{InterEQM}
\end{eqnarray}

In order to fully describe the electron transport problem, we should record
the number of electrons arriving at the drain electrode, which emitted from the
source electrode and passing through the QD molecule. We follow Li and
co-authors \cite{Li02,WangSK} and introduce the Hilbert subspace $B^{\left(
n\right) }$ $\left( n=1,2,...\right) $ corresponding to $n$ electrons
arriving at the drain electrode, which is spanned by the product of all
many-particle states of the two isolated electrodes, and formally denoted as
$B^{\left( n\right) }\equiv $ span$\left\{ \left\vert \Psi _{L}\right\rangle
^{\left( n\right) }\otimes \left\vert \Psi _{R}\right\rangle ^{\left(
n\right) }\right\} $. Then, the entire Hilbert space of the two electrodes
can be expressed as $B=\oplus _{n}B^{\left( n\right) }$. With this
classification of the electrode states, the average over states in the
entire Hilbert space $B$ in Eq. (\ref{InterEQM}) should be replaced with the
states in the subspace $B^{\left( n\right) }$, and leading to a conditional
TCL master equation
\begin{eqnarray}
&&\frac{\partial }{\partial t}\rho _{I,S}^{\left( n\right) }\left( t\right)
\notag \\
&=&-\sum_{\alpha \mu \nu }\int_{-\infty }^{t}dt_{1}\text{tr}_{B^{\left(
n\right) }}\left[ \rho _{I,S}\left( t\right) \otimes \rho _{B}f_{\alpha \nu
}^{\dag }\left( t_{1}\right) d_{\nu }\left( t_{1}\right) d_{\mu }^{\dag
}\left( t\right) f_{\alpha \mu }\left( t\right) \right]  \notag \\
&&-\sum_{\alpha \mu \nu }\int_{-\infty }^{t}dt_{1}\text{tr}_{B^{\left(
n\right) }}\left[ d_{\mu }^{\dag }\left( t\right) f_{\alpha \mu }\left(
t\right) f_{\alpha \nu }^{\dag }\left( t_{1}\right) d_{\nu }\left(
t_{1}\right) \rho _{I,S}\left( t\right) \otimes \rho _{B}\right]  \notag \\
&&+\sum_{\alpha \mu \nu }\int_{-\infty }^{t}dt_{1}\text{tr}_{B^{\left(
n\right) }}\left[ f_{\alpha \nu }^{\dag }\left( t_{1}\right) d_{\nu }\left(
t_{1}\right) \rho _{I,S}\left( t\right) \otimes \rho _{B}d_{\mu }^{\dag
}\left( t\right) f_{\alpha \mu }\left( t\right) \right]  \notag \\
&&+\sum_{\alpha \mu \nu }\int_{-\infty }^{t}dt_{1}\text{tr}_{B^{\left(
n\right) }}\left[ d_{\mu }^{\dag }\left( t\right) f_{\alpha \mu }\left(
t\right) \rho _{I,S}\left( t\right) \otimes \rho _{B}f_{\alpha \nu }^{\dag
}\left( t_{1}\right) d_{\nu }\left( t_{1}\right) \right] +H.c..
\label{ndensity}
\end{eqnarray}

To proceed, two physical considerations are further implemented. (i) Instead
of the conventional Born approximation for the entire density matrix $\rho
_{T}\left( t\right) \simeq \rho \left( t\right) \otimes \rho _{B}$, the
ansatz $\rho ^{I}\left( t\right) \simeq \rho ^{\left( n\right) }\left(
t\right) \otimes \rho _{B}^{\left( n\right) }$ is proposed, where $\rho
_{B}^{\left( n\right) }$ being the electrode density operator associated
with $n$ electrons arriving at the drain electrode. With this ansatz for the
entire density operator, tracing over the subspace $B^{\left( n\right) }$,
the Eq. (\ref{ndensity}) can be rewritten as
\begin{eqnarray}
&&\frac{\partial }{\partial t}\rho _{I,S}^{\left( n\right) }\left( t\right)
\notag \\
&=&-\sum_{\alpha \mu \nu }\int_{-\infty }^{t}dt_{1}\text{tr}_{B^{\left(
n\right) }}\left[ f_{\alpha \nu }^{\dag }\left( t_{1}\right) f_{\alpha \mu
}\left( t\right) \rho _{B}\right] \rho _{I,S}^{\left( n\right) }\left(
t\right) d_{\nu }\left( t_{1}\right) d_{\mu }^{\dag }\left( t\right)   \notag
\\
&&-\sum_{\alpha \mu \nu }\int_{-\infty }^{t}dt_{1}\text{tr}_{B^{\left(
n\right) }}\left[ f_{\alpha \mu }\left( t\right) f_{\alpha \nu }^{\dag
}\left( t_{1}\right) \rho _{B}\right] d_{\mu }^{\dag }\left( t\right) d_{\nu
}\left( t_{1}\right) \rho _{I,S}^{\left( n\right) }\left( t\right)   \notag
\\
&&+\sum_{\mu \nu }\int_{-\infty }^{t}dt_{1}\text{tr}_{B^{\left( n\right) }}%
\left[ f_{L\mu }\left( t\right) f_{L\nu }^{\dag }\left( t_{1}\right) \rho
_{B}\right] d_{\nu }\left( t_{1}\right) \rho _{I,S}^{\left( n\right) }\left(
t\right) d_{\mu }^{\dag }\left( t\right)   \notag \\
&&+\sum_{\mu \nu }\int_{-\infty }^{t}dt_{1}\text{tr}_{B^{\left( n\right) }}%
\left[ f_{R\mu }\left( t\right) f_{R\nu }^{\dag }\left( t_{1}\right) \rho
_{B}\right] d_{\nu }\left( t_{1}\right) \rho _{I,S}^{\left( n-1\right)
}\left( t\right) d_{\mu }^{\dag }\left( t\right)   \notag \\
&&+\sum_{\mu \nu }\int_{-\infty }^{t}dt_{1}\text{tr}_{B^{\left( n\right) }}%
\left[ f_{L\nu }^{\dag }\left( t_{1}\right) f_{L\mu }\left( t\right) \rho
_{B}\right] d_{\mu }^{\dag }\left( t\right) \rho _{I,S}^{\left( n\right)
}\left( t\right) d_{\nu }\left( t_{1}\right)   \notag \\
&&+\sum_{\mu \nu }\int_{-\infty }^{t}dt_{1}\text{tr}_{B^{\left( n\right) }}%
\left[ f_{R\nu }^{\dag }\left( t_{1}\right) f_{R\mu }\left( t\right) \rho
_{B}\right] d_{\mu }^{\dag }\left( t\right) \rho _{I,S}^{\left( n+1\right)
}\left( t\right) d_{\nu }\left( t_{1}\right) +H.c..  \label{ndensitymodified}
\end{eqnarray}%
Here we have used the orthogonality between the states in different
subspaces. (ii) The extra electrons arriving at the drain electrode will
flow back into the source electrode via the external closed transport
circuit. Moreover, the rapid relaxation processes in the electrodes
will bring the electrodes to the local thermal equilibrium states quickly,
which are determined by the chemical potentials. Consequently, after the
procedure done in Eq. (\ref{ndensitymodified}), the electrode density
matrices $\rho _{B}^{\left( n\right) }$ and $\rho _{B}^{\left( n\pm 1\right)
}$ should be replaced by $\rho _{B}^{\left( 0\right) }$. In the Schr\"{o}%
dinger representation, the Eq. (\ref{ndensitymodified}) can be expressed as
\begin{eqnarray}
&&\frac{\partial }{\partial t}\rho _{S}^{\left( n\right) }\left( t\right)
\notag \\
&=&-i\left[ H_{S},\rho _{S}^{\left( n\right) }\left( t\right) \right]
\notag \\
&&-\sum_{\alpha \mu \nu }\int_{-\infty }^{t}dt_{1}C_{\alpha \nu \mu
}^{\left( +\right) }\left( t_{1}-t\right) \rho _{S}^{\left( n\right) }\left(
t\right) e^{-iH_{S}\left( t-t_{1}\right) }d_{\nu }e^{iH_{S}\left(
t-t_{1}\right) }d_{\mu }^{\dag }  \notag \\
&&-\sum_{\alpha \mu \nu }\int_{-\infty }^{t}dt_{1}C_{\alpha \mu \nu
}^{\left( -\right) }\left( t-t_{1}\right) d_{\mu }^{\dag }e^{-iH_{S}\left(
t-t_{1}\right) }d_{\nu }e^{iH_{S}\left( t-t_{1}\right) }\rho _{S}^{\left(
n\right) }\left( t\right)   \notag \\
&&+\sum_{\mu \nu }\int_{-\infty }^{t}dt_{1}C_{L\mu \nu }^{\left( -\right)
}\left( t-t_{1}\right) e^{-iH_{S}\left( t-t_{1}\right) }d_{\nu
}e^{iH_{S}\left( t-t_{1}\right) }\rho _{S}^{\left( n\right) }\left( t\right)
d_{\mu }^{\dag }  \notag \\
&&+\sum_{\mu \nu }\int_{-\infty }^{t}dt_{1}C_{R\mu \nu }^{\left( -\right)
}\left( t-t_{1}\right) e^{-iH_{S}\left( t-t_{1}\right) }d_{\nu
}e^{iH_{S}\left( t-t_{1}\right) }\rho _{S}^{\left( n-1\right) }\left(
t\right) d_{\mu }^{\dag }  \notag \\
&&+\sum_{\mu \nu }\int_{-\infty }^{t}dt_{1}C_{L\nu \mu }^{\left( +\right)
}\left( t_{1}-t\right) d_{\mu }^{\dag }\rho _{S}^{\left( n\right) }\left(
t\right) e^{-iH_{S}\left( t-t_{1}\right) }d_{\nu }e^{iH_{S}\left(
t-t_{1}\right) }  \notag \\
&&+\sum_{\mu \nu }\int_{-\infty }^{t}dt_{1}C_{R\nu \mu }^{\left( +\right)
}\left( t_{1}-t\right) d_{\mu }^{\dag }\rho _{S}^{\left( n+1\right) }\left(
t\right) e^{-iH_{S}\left( t-t_{1}\right) }d_{\nu }e^{iH_{S}\left(
t-t_{1}\right) }+H.c..  \label{nQME}
\end{eqnarray}%
where the correlation function are defined as
\begin{align}
C_{\alpha \mu \nu }^{\left( +\right) }\left( t-t_{1}\right) & =\text{tr}_{R}%
\left[ f_{\alpha \mu }^{\dag }\left( t\right) f_{\alpha \nu }\left(
t_{1}\right) \rho _{B}\right] =\left\langle f_{\alpha \mu }^{\dagger }\left(
t\right) f_{\alpha \nu }\left( t_{1}\right) \right\rangle ,  \notag \\
C_{\alpha \mu \nu }^{\left( -\right) }\left( t-t_{1}\right) & =\text{tr}_{R}%
\left[ f_{\alpha \mu }\left( t\right) f_{\alpha \nu }^{\dag }\left(
t_{1}\right) \rho _{B}\right] =\left\langle f_{\alpha \mu }\left( t\right)
f_{\alpha \nu }^{\dagger }\left( t_{1}\right) \right\rangle .
\label{Correfun}
\end{align}%
Introducing the following super-operators
\begin{align}
A_{\alpha \mu }^{\left( +\right) }\left( t\right) & =\sum_{\nu
}\int_{-\infty }^{t}dt_{1}C_{\alpha \nu \mu }^{\left( +\right) }\left(
t_{1}-t\right) e^{-iH_{S}\left( t-t_{1}\right) }d_{\nu }e^{iH_{S}\left(
t-t_{1}\right) },  \notag \\
A_{\alpha \mu }^{\left( -\right) }\left( t\right) & =\sum_{\nu
}\int_{-\infty }^{t}dt_{1}C_{\alpha \mu \nu }^{\left( -\right) }\left(
t-t_{1}\right) e^{-iH_{S}\left( t-t_{1}\right) }d_{\nu }e^{iH_{S}\left(
t-t_{1}\right) },  \label{superoperator}
\end{align}%
then, the Eq. (\ref{nQME}) can be rewritten as a compact form
\begin{eqnarray}
&&\frac{\partial }{\partial t}\rho _{S}^{\left( n\right) }\left( t\right)
\notag \\
&=&-i\left[ H_{S},\rho _{S}^{\left( n\right) }\left( t\right) \right]
\notag \\
&&-\sum_{\mu }\left\{ \rho _{S}^{\left( n\right) }\left( t\right) A_{\mu
}^{\left( +\right) }\left( t\right) d_{\mu }^{\dag }+d_{\mu }^{\dag }A_{\mu
}^{\left( -\right) }\left( t\right) \rho _{S}^{\left( n\right) }\left(
t\right) \right.   \notag \\
&&-A_{L\mu }^{\left( -\right) }\left( t\right) \rho _{S}^{\left( n\right)
}\left( t\right) d_{\mu }^{\dag }-A_{R\mu }^{\left( -\right) }\left(
t\right) \rho _{S}^{\left( n-1\right) }\left( t\right) d_{\mu }^{\dag }
\notag \\
&&\left. -d_{\mu }^{\dag }\rho _{S}^{\left( n\right) }\left( t\right)
A_{L\mu }^{\left( +\right) }\left( t\right) -d_{\mu }^{\dag }\rho
_{S}^{\left( n+1\right) }\left( t\right) A_{R\mu }^{\left( +\right) }\left(
t\right) +H.c.\right\} .  \label{nQMEfinal}
\end{eqnarray}%
where $A_{\mu }^{\left( \pm \right) }\left( t\right) =\sum_{\alpha
}A_{\alpha \mu }^{\left( \pm \right) }\left( t\right) $. The above equation
is the starting point of the non-Markovian FCS calculation.

\subsection*{Non-Markovian full counting statistics}

In this subsection, we outline the procedure to calculate the non-Markovian
FCS based on Eq. (\ref{nQMEfinal}). The FCS can be obtained from the
cumulant generating function (CGF) $F\left( \chi \right) $ which related to
the probability distribution $P\left( n,t\right) $ by\cite{WangSK,Bagrets} $%
e^{-F\left( \chi \right) }=\sum_{n}P\left( n,t\right) e^{in\chi }$, where $%
\chi $ is the counting field. The CGF $F\left( \chi \right) $ connects with
the particle-number-resolved density matrix $\rho ^{\left( n\right) }\left(
t\right) $ by defining $S\left( \chi ,t\right) =\sum_{n}\rho ^{\left(
n\right) }\left( t\right) e^{in\chi }$. Evidently, we have $e^{-F\left( \chi
\right) }=$Tr$\left[ S\left( \chi ,t\right) \right] $, where the trace is
over the eigenstates of the QD molecule system. Since Eq. (\ref{nQMEfinal})
has the following form $\dot{\rho}^{\left( n\right) }=A\rho ^{\left(
n\right) }+C_{1}\rho ^{\left( n+1\right) }+D_{1}\rho ^{\left( n-1\right) }$,
then, $S\left( \chi ,t\right) $ satisfies $\dot{S}=AS+e^{-i\chi }CS+e^{i\chi
}DS\equiv L_{\chi }S$, where $S$ is a column matrix, and $A$, $C$ and $D$
are three square matrices. The specific form of $L_{\chi }$ can be obtained
by performing a discrete Fourier transformation to the matrix element of Eq.
(\ref{nQMEfinal}). In the low frequency limit, the counting time, namely, the time of
measurement is much longer than the time of tunneling through the QD
molecule system. In this case, $F\left( \chi \right) $ is given by\cite%
{Flindt01,Bagrets,Flindt02,Flindt03,Kieblich01,Groth} $F\left( \chi \right)
=-\lambda _{1}\left( \chi \right) t$, where $\lambda _{1}\left( \chi \right)
$ is the eigenvalue of $L_{\chi }$ which goes to zero for $\chi \rightarrow 0
$. According to the definition of the cumulants one can express $\lambda
_{1}\left( \chi \right) $\ as $\lambda _{1}\left( \chi \right)
=\sum_{k=1}^{\infty }\frac{C_{k}}{t}\frac{\left( i\chi \right) ^{k}}{k!}$.
The low order cumulants can be calculated by the Rayleigh--Schr\"{o}dinger
perturbation theory in the counting parameter $\chi $. In order to calculate
the first four current cumulants we expand $L_{\chi }$ to four order in $%
\chi $
\begin{equation}
L_{\chi }=L_{0}+L_{1}\chi +\frac{1}{2!}L_{2}\chi ^{2}+\frac{1}{3!}L_{3}\chi
^{3}+\frac{1}{4!}L_{4}\chi ^{4}+\cdots .  \label{matirxL}
\end{equation}%
and define the two projectors \cite{Flindt01,Flindt02,Flindt03,XueJAP11} $%
P=P^{2}=\left\vert \left. 0\right\rangle \right\rangle \left\langle
\left\langle \tilde{0}\right. \right\vert $ and $Q=Q^{2}=1-P$, obeying the
relations $PL_{0}=L_{0}P=0$ and $QL_{0}=L_{0}Q=L_{0}$. Here, $\left\vert
\left. 0\right\rangle \right\rangle $
is the right eigenvector of $L_{0}$, i.e., $L_{0}\left\vert \left.
0\right\rangle \right\rangle =0$, and $\left\langle \left\langle \tilde{0}%
\right. \right\vert \equiv \hat{1}$ is the corresponding left eigenvector.
In view of $L_{0}$ being regular, we also introduce the pseudoinverse
according to $R=QL_{0}^{-1}Q$, which is well-defined due to the inversion
being performed only in the subspace spanned by $Q$. After a careful
calculation, $\lambda _{1}\left( \chi \right) $ is given by
\begin{align}
\lambda _{1}\left( \chi \right) & =\left\langle \left\langle \tilde{0}%
\right. \right\vert L_{1}\left\vert \left. 0\right\rangle \right\rangle \chi
\notag \\
& +\frac{1}{2!}\left[ \left\langle \left\langle \tilde{0}\right. \right\vert
L_{2}\left\vert \left. 0\right\rangle \right\rangle -2\left\langle
\left\langle \tilde{0}\right. \right\vert L_{1}RL_{1}\left\vert \left.
0\right\rangle \right\rangle \right] \chi ^{2}  \notag \\
& +\frac{1}{3!}\left[ \left\langle \left\langle \tilde{0}\right. \right\vert
L_{3}\left\vert \left. 0\right\rangle \right\rangle -3\left\langle
\left\langle \tilde{0}\right. \right\vert \left(
L_{2}RL_{1}+L_{1}RL_{2}\right) \left\vert \left. 0\right\rangle
\right\rangle \right.   \notag \\
& \left. -6\left\langle \left\langle \tilde{0}\right. \right\vert
L_{1}R\left( RL_{1}P-L_{1}R\right) L_{1}\left\vert \left. 0\right\rangle
\right\rangle \right] \chi ^{3}+  \notag \\
& +\frac{1}{4!}\left[ \left\langle \left\langle \tilde{0}\right. \right\vert
L_{4}\left\vert \left. 0\right\rangle \right\rangle -6\left\langle
\left\langle \tilde{0}\right. \right\vert L_{2}RL_{2}\left\vert \left.
0\right\rangle \right\rangle \right.   \notag \\
& -4\left\langle \left\langle \tilde{0}\right. \right\vert \left(
L_{3}RL_{1}+L_{1}RL_{3}\right) \left\vert \left. 0\right\rangle
\right\rangle   \notag \\
& -12\left\langle \left\langle \tilde{0}\right. \right\vert L_{2}R\left(
RL_{1}P-L_{1}R\right) L_{1}\left\vert \left. 0\right\rangle \right\rangle
\notag \\
& -12\left\langle \left\langle \tilde{0}\right. \right\vert L_{1}R\left(
RL_{2}P-L_{2}R\right) L_{1}\left\vert \left. 0\right\rangle \right\rangle
\notag \\
& -12\left\langle \left\langle \tilde{0}\right. \right\vert L_{1}R\left(
RL_{1}P-L_{1}R\right) L_{2}\left\vert \left. 0\right\rangle \right\rangle
\notag \\
& -24\left\langle \left\langle \tilde{0}\right. \right\vert L_{1}R\left(
R^{2}L_{1}PL_{1}P-RL_{1}PL_{1}R-L_{1}R^{2}L_{1}P\right.   \notag \\
& \left. \left. -RL_{1}RL_{1}P+L_{1}RL_{1}R\right) L_{1}\left\vert \left.
0\right\rangle \right\rangle \right] \chi ^{4}+\cdots .  \label{matrixLambda}
\end{align}%
From Eq. (\ref{matrixLambda}) we can identify the first
four current cumulants:
\begin{equation}
C_{1}/t=\left\langle \left\langle \tilde{0}\right. \right\vert
L_{1}\left\vert \left. 0\right\rangle \right\rangle /i,  \label{current}
\end{equation}%
\begin{equation}
C_{2}/t=\left[ \left\langle \left\langle \tilde{0}\right. \right\vert
L_{2}\left\vert \left. 0\right\rangle \right\rangle -2\left\langle
\left\langle \tilde{0}\right. \right\vert L_{1}RL_{1}\left\vert \left.
0\right\rangle \right\rangle \right] /i^{2},  \label{shot noise}
\end{equation}%
\begin{align}
& C_{3}/t=\left[ \left\langle \left\langle \tilde{0}\right. \right\vert
L_{3}\left\vert \left. 0\right\rangle \right\rangle -3\left\langle
\left\langle \tilde{0}\right. \right\vert \left(
L_{2}RL_{1}+L_{1}RL_{2}\right) \left\vert \left. 0\right\rangle
\right\rangle \right.   \notag \\
& \left. -6\left\langle \left\langle \tilde{0}\right. \right\vert
L_{1}R\left( RL_{1}P-L_{1}R\right) L_{1}\left\vert \left. 0\right\rangle
\right\rangle \right] /i^{3}.  \label{skewness}
\end{align}%
\begin{align}
& C_{4}/t=\left[ \left\langle \left\langle \tilde{0}\right. \right\vert
L_{4}\left\vert \left. 0\right\rangle \right\rangle -6\left\langle
\left\langle \tilde{0}\right. \right\vert L_{2}RL_{2}\left\vert \left.
0\right\rangle \right\rangle \right.   \notag \\
& -4\left\langle \left\langle \tilde{0}\right. \right\vert \left(
L_{3}RL_{1}+L_{1}RL_{3}\right) \left\vert \left. 0\right\rangle
\right\rangle   \notag \\
& -12\left\langle \left\langle \tilde{0}\right. \right\vert L_{2}R\left(
RL_{1}P-L_{1}R\right) L_{1}\left\vert \left. 0\right\rangle \right\rangle
\notag \\
& -12\left\langle \left\langle \tilde{0}\right. \right\vert L_{1}R\left(
RL_{2}P-L_{2}R\right) L_{1}\left\vert \left. 0\right\rangle \right\rangle
\notag \\
& -12\left\langle \left\langle \tilde{0}\right. \right\vert L_{1}R\left(
RL_{1}P-L_{1}R\right) L_{2}\left\vert \left. 0\right\rangle \right\rangle
\notag \\
& -24\left\langle \left\langle \tilde{0}\right. \right\vert L_{1}R\left(
R^{2}L_{1}PL_{1}P-RL_{1}PL_{1}R-L_{1}R^{2}L_{1}P\right.   \notag \\
& \left. \left. -RL_{1}RL_{1}P+L_{1}RL_{1}R\right) L_{1}\left\vert \left.
0\right\rangle \right\rangle \right] /i^{4}.  \label{kurtosis}
\end{align}%
Here, it is important to emphasize that the first four cumulants $%
C_{k}$ are directly related to the transport characteristics. For example,
the first-order cumulant (the peak position of the distribution of
transferred-electron number) $C_{1}=\bar{n}$ gives the average current $%
\left\langle I\right\rangle =eC_{1}/t$. The zero-frequency shot noise is
related to the second-order cumulant (the peak-width of the distribution) $%
S=2e^{2}C_{2}/t=2e^{2}\left( \overline{n^{2}}-\bar{n}^{2}\right) /t$. The
third-order cumulant $C_{3}=\overline{\left( n-\bar{n}\right) ^{3}}$ and four-order cumulant $C_{4}=\overline{\left( n-\bar{n}\right) ^{4}}%
-3\overline{\left( n-\bar{n}\right) ^{2}}^{2}$ characterize, respectively,
the skewness and kurtosis of the distribution. Here, $\overline{\left(
\cdots \right) }=\sum_{n}\left( \cdots \right) P\left( n,t\right) $. In
general, the shot noise, skewness and kurtosis are represented by the Fano
factor $F_{2}=C_{2}/C_{1}$, $F_{3}=C_{3}/C_{1}$ and $F_{4}=C_{4}/C_{1}$,
respectively.

\section*{Acknowledgments}
This work was supported by the NKBRSFC under grants Nos. 2011CB921502,
2012CB821305, NSFC under grants Nos. 11204203, 61405138, 11004124, 11275118, 61227902,
61378017, 11434015, SKLQOQOD under grants No. KF201403, SPRPCAS under grants
No. XDB01020300.

\section*{Author Contributions}
H. B. X. conceived the idea and designed the research and performed calculations.
H. J. J, J. Q. L. and W. M. L. contributed to the analysis and interpretation of the results and prepared the manuscript.

\section*{Competing Interests}
The authors declare no competing financial interests.

\section*{Correspondence}
Correspondence and requests for materials should be addressed to Hai-Bin Xue or Wu-Ming Liu.

\clearpage
\newpage

\begin{figure*}[t]
\centerline{\includegraphics[height=12cm,width=16cm]{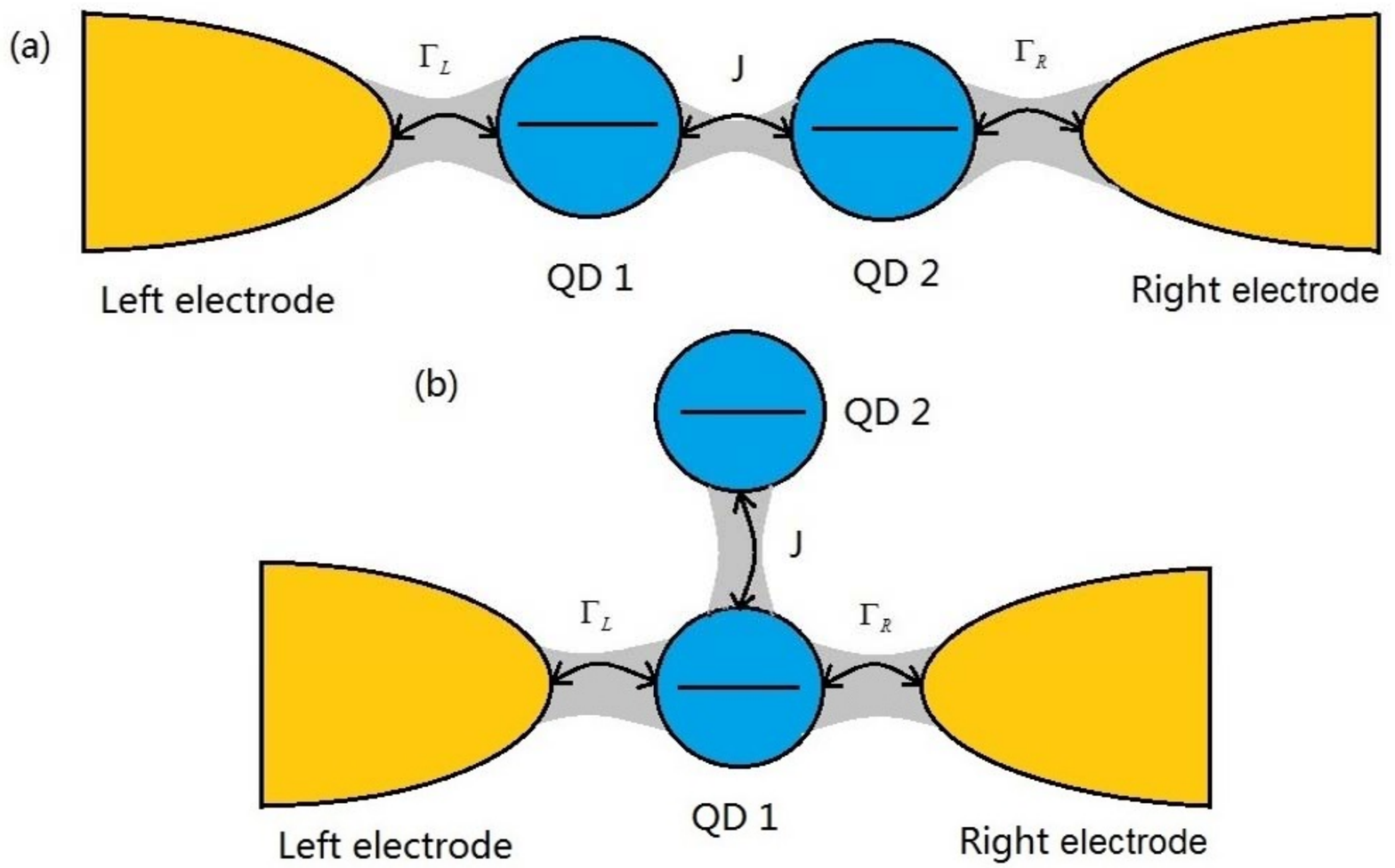}}
\caption{Schematic of the two single-level QD molecules weakly
coupled to two electrodes, (a) serially
coupled double QDs, (b) side-coupled double QDs. Here, the two QD molecules
possess high quantum coherence in the case of $\Delta \ll k_{B}T$ (%
$\Delta $ being the singly-occupied eigenenergy separation, $k_{B}$ the
Boltzmann constant, $T$ the temperature of the QDs system). The
hopping coupling between the two QDs, and the strength of coupling
between the QDs system and the electrode $\alpha$, are
characterized by $J$ and $\Gamma _{\alpha}$, respectively.} \label{fig1}
\end{figure*}

\begin{figure*}[t]
\centerline{\includegraphics[height=12cm,width=16cm]{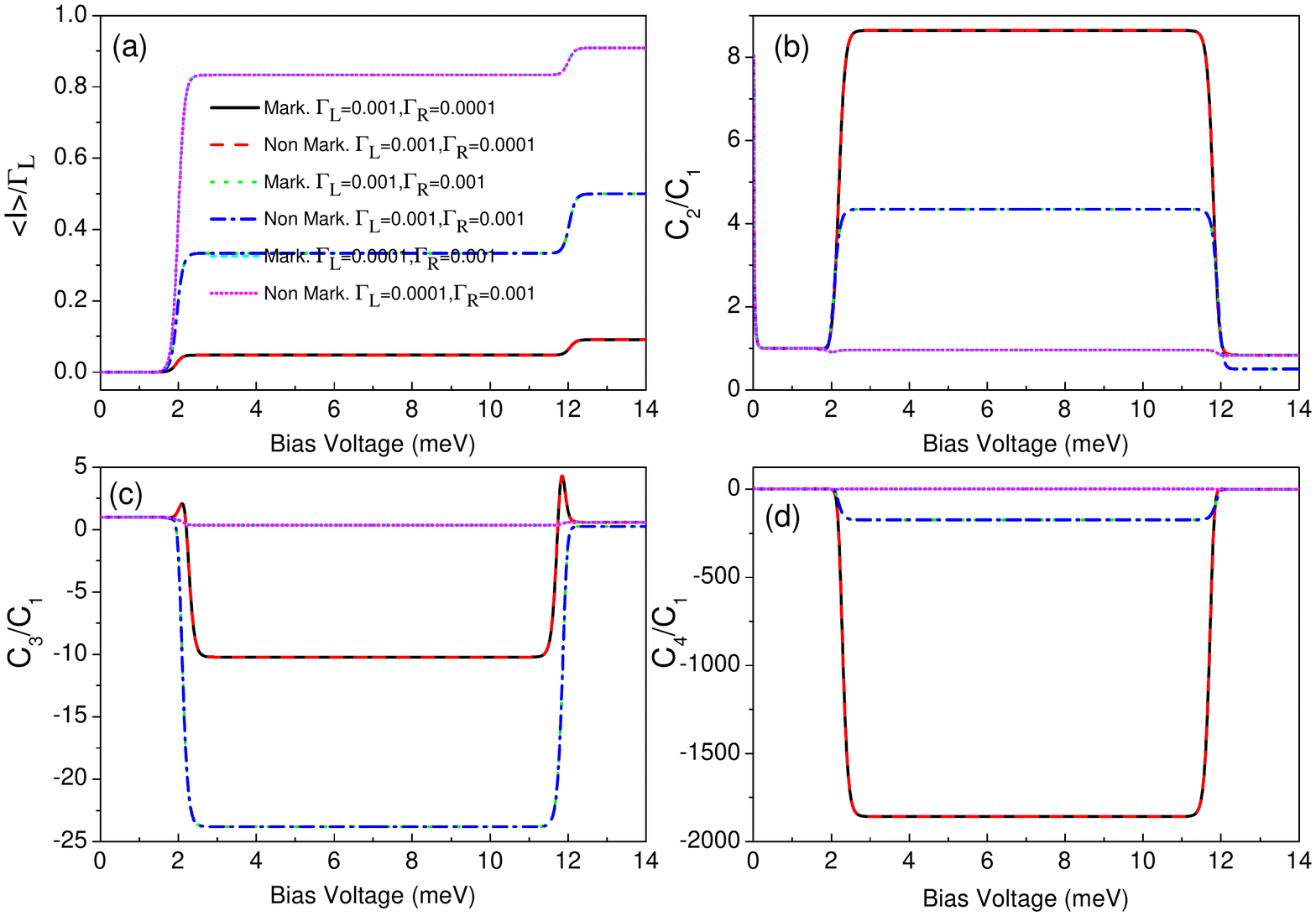}}
\caption{The average current ($\left\langle I\right\rangle$),
shot noise ($C_{2}/C_{1}$), skewness ($C_{3}/C_{1}$) and kurtosis ($%
C_{4}/C_{1}$) versus bias voltage for the Morkovian and the non-Markovian
case at different coupling of the single QD with two ferromagnetic
electrodes, respectively. Here, $C_{k}$ is the zero-frequency ${k}$-order
cumulant of current fluctuations. The non-Markovian effect has no influence
on the first four current cumulants of the considered single QD. The single
QD system parameters: $\epsilon_{\uparrow }=\epsilon _{\downarrow }=1$, $U=5$, $p=0.9$ and $k_{B}T=0.04$.} \label{fig2}
\end{figure*}

\begin{figure*}[t]
\centerline{\includegraphics[height=12cm,width=16cm]{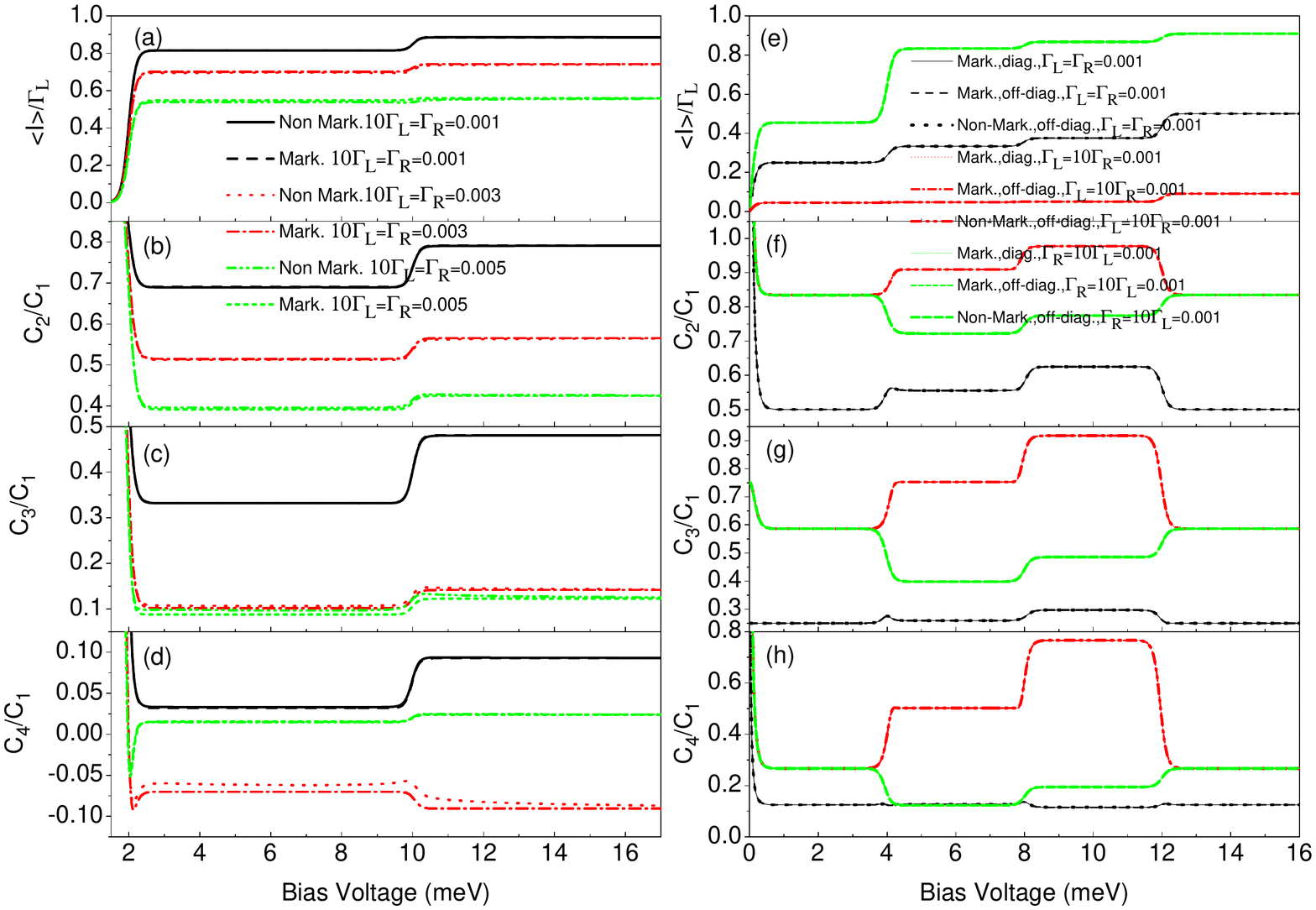}}
\caption{(a)-(d) The average current ($\left\langle
I\right\rangle$), shot noise ($C_{2}/C_{1}$), skewness ($C_{3}/C_{1}$) and
kurtosis ($C_{4}/C_{1}$) versus bias voltage for the Morkovian and the
non-Markovian case at different values of the QD-2-electrode coupling $%
\Gamma _{R}$ with $\Gamma _{L}/\Gamma _{R}=0.1$. Here, $C_{k}$ is the
zero-frequency ${k}$-order cumulant of current fluctuations. The non-Markovian
effect in the $\Gamma _{L}/\Gamma _{R}=0.1$ case has a
weak influence on the the first four current cumulants. The serially
coupled double QDs system parameters: $\epsilon _{1}=\epsilon _{2}=1$, $%
J=0.001$, $U=4$ and $k_{B}T=0.05$.
(e)-(h) The average current ($\left\langle I\right\rangle$), shot noise ($C_{2}/C_{1}$), skewness ($%
C_{3}/C_{1}$) and kurtosis ($C_{4}/C_{1}$) versus bias voltage for different
coupling of the serially coupled double QDs system with two metallic
electrodes. Here three cases are considered, namely, (1) the Markovian and the
diagonal elements of the reduced density matrix, (2) the Markovian and the
off-diagonal elements of the reduced density matrix, (3) the non-Markovian
and the off-diagonal elements of the reduced density matrix. The non-Markovian
effect has a very weak influence on the first four current
cumulants in the serially coupled double QD system with a relatively small quantum coherence. The serially
coupled double QDs system parameters: $\epsilon _{1}=\epsilon _{2}=1$, $J=1$%
, $U=4$ and $k_{B}T=0.05$.} \label{fig3}
\end{figure*}

\begin{figure*}[t]
\centerline{\includegraphics[height=12cm,width=16cm]{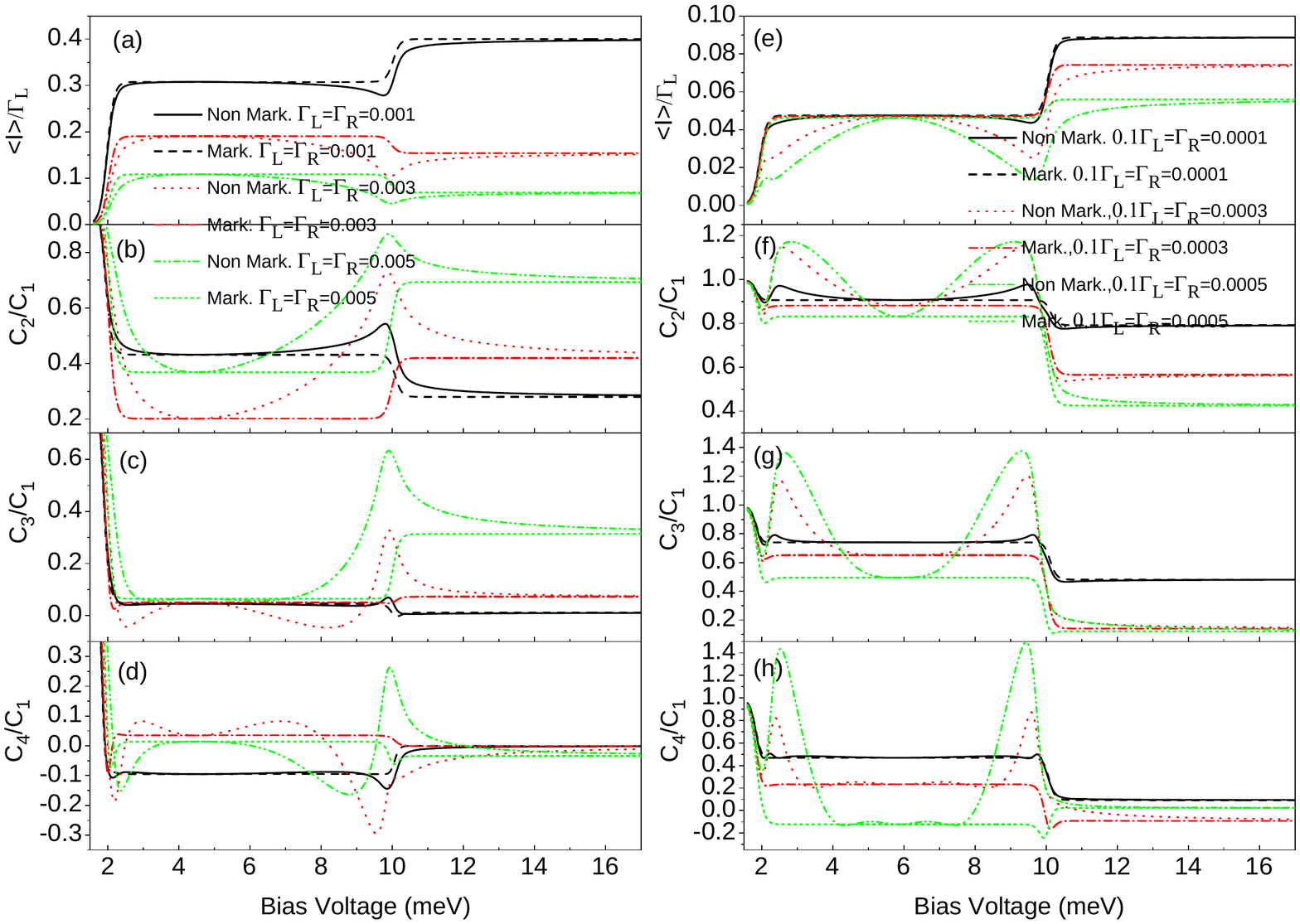}}
\caption{The average current ($\left\langle I\right\rangle$),
shot noise ($C_{2}/C_{1}$), skewness ($C_{3}/C_{1}$) and kurtosis ($%
C_{4}/C_{1}$) versus bias voltage for the Morkovian and the non-Markovian
case at different values of the QD-2-electrode coupling $\Gamma _{R}$.
(a)-(d) for $\Gamma _{L}/\Gamma _{R}=1$, (e)-(h) for $\Gamma _{L}/\Gamma
_{R}=10$. Here, $C_{k}$ is the zero-frequency ${k}$-order cumulant of
current fluctuations. The non-Markovian effect in the $\Gamma _{L}/\Gamma _{R}\geq 1$ case has a
significant impact on the first four cumulants of transport current. The serially coupled double QDs system parameters: $%
\epsilon _{1}=\epsilon _{2}=1$, $J=0.001$, $U=4$ and $k_{B}T=0.05$.} \label{fig4}
\end{figure*}

\begin{figure*}[t]
\centerline{\includegraphics[height=12cm,width=16cm]{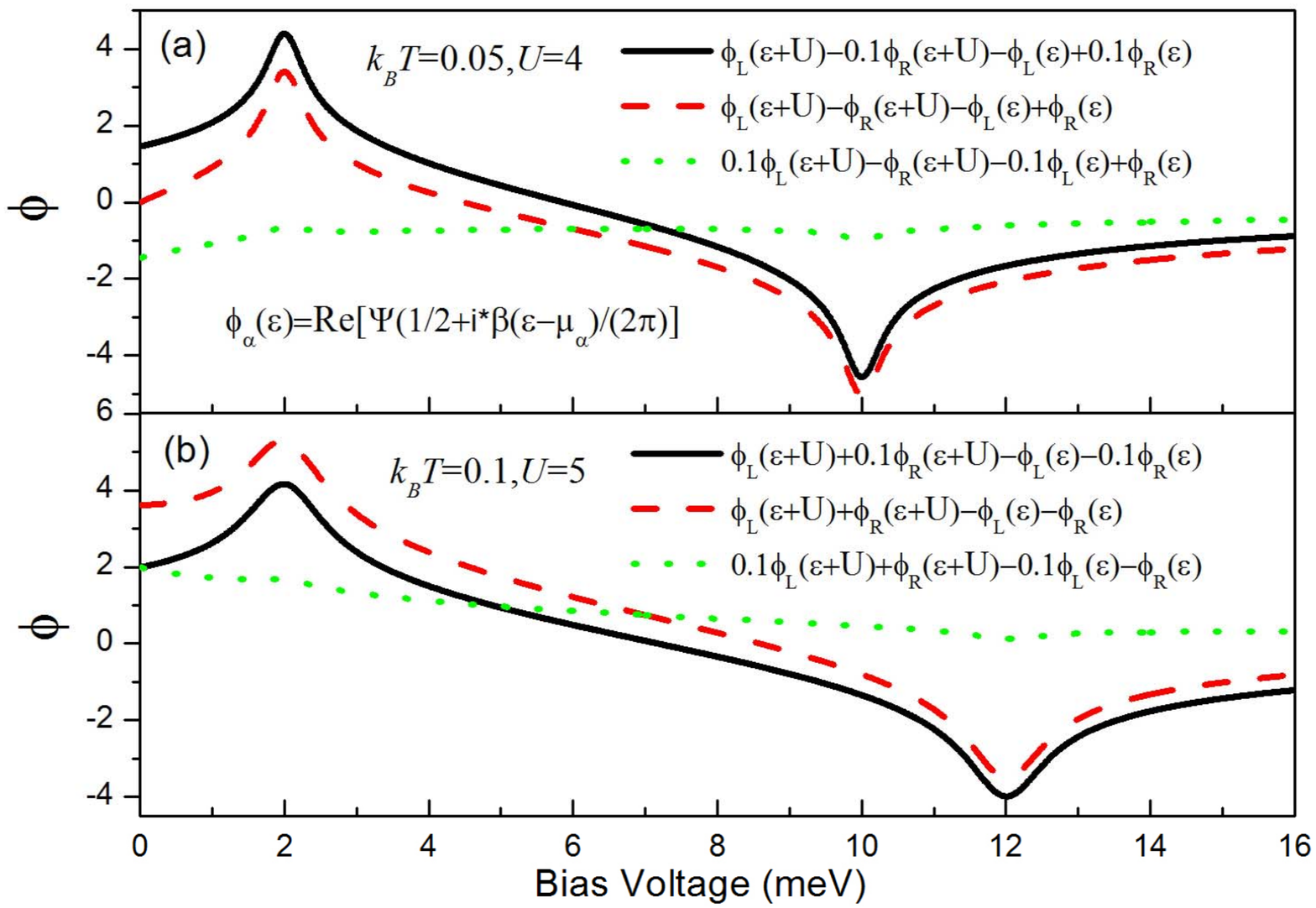}}
\caption{(a) The functions $\Phi _{L}-0.1\Phi _{R}$ $\left(
\Gamma _{R}=0.1\Gamma _{L}\right) $, $\Phi _{L}-\Phi _{R}$ $\left( \Gamma
_{R}=\Gamma _{L}\right) $ and $0.1\Phi _{L}-\Phi _{R}$ $\left( \Gamma
_{L}=0.1\Gamma _{R}\right) $ as a function of bias voltage with $U=4$ and $%
k_{B}T=0.05$. (b) The functions $\Phi _{L}+0.1\Phi _{R}$ $\left( \Gamma
_{R}=0.1\Gamma _{L}\right) $, $\Phi _{L}+\Phi _{R}$ $\left( \Gamma
_{R}=\Gamma _{L}\right) $ and $0.1\Phi _{L}+\Phi _{R}$ $\left( \Gamma
_{L}=0.1\Gamma _{R}\right) $ as a function of bias voltage with $U=5$ and $%
k_{B}T=0.1$. Here, $\Phi _{\alpha }=\phi _{\alpha }\left( \epsilon +U\right)
-\phi _{\alpha }\left( \epsilon \right) $, $\phi _{\alpha }\left( \epsilon
\right) =\operatorname{Re}\Psi \left[ \frac{1}{2}+\frac{i\left( \epsilon -\mu
_{L}\right) }{2\pi k_{B}T}\right] $ and $\Psi $ is the digamma function. The variation of the value of the above
mentioned function is responsible for whether the non-Markovian effect has
a remarkable influence on the first four cumulants of transport current.} \label{fig5}
\end{figure*}

\begin{figure*}[t]
\centerline{\includegraphics[height=12cm,width=16cm]{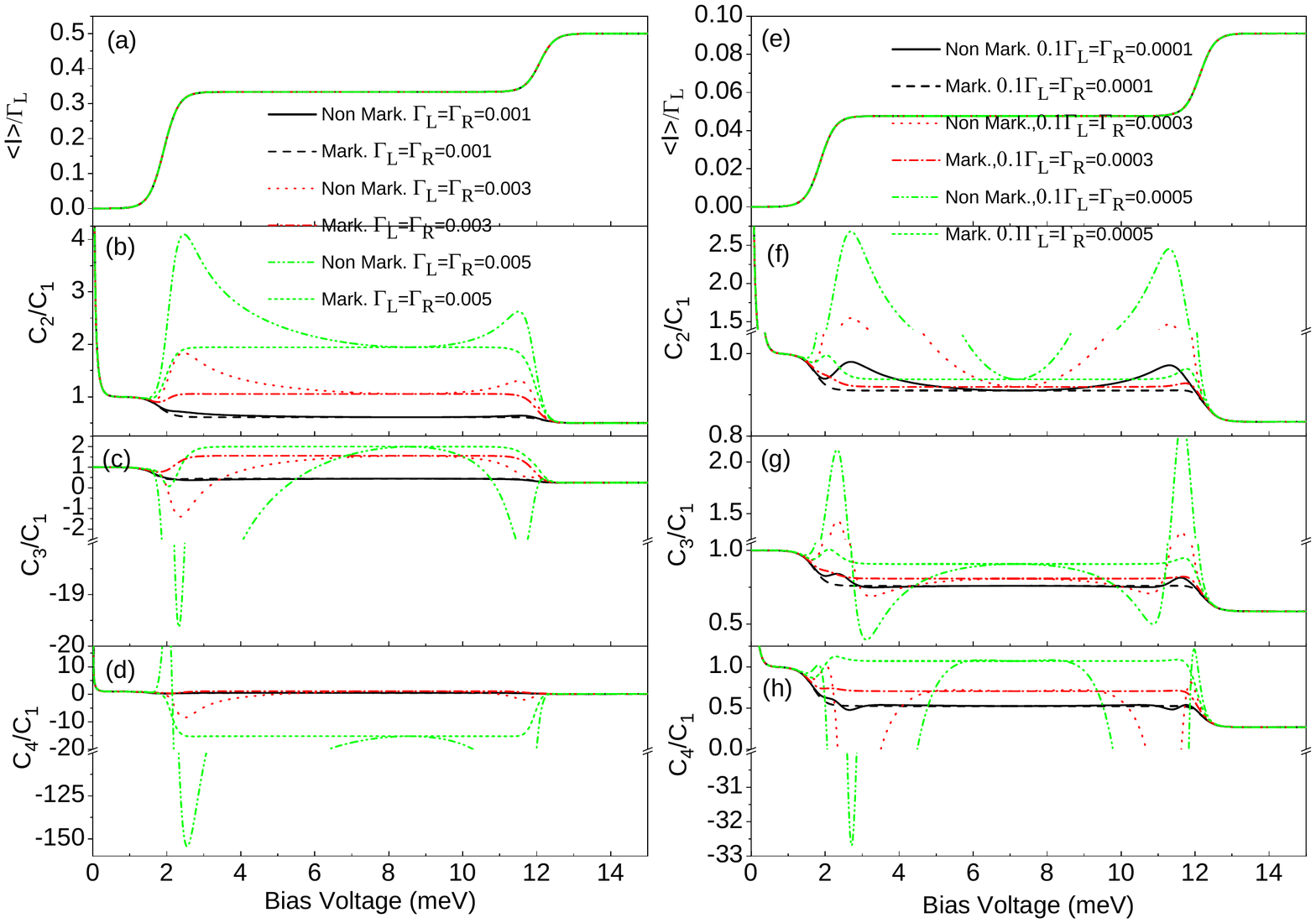}}
\caption{The average current ($\left\langle I\right\rangle$),
shot noise ($C_{2}/C_{1}$), skewness ($C_{3}/C_{1}$) and kurtosis ($%
C_{4}/C_{1}$) versus bias voltage for the Morkovian and the non-Markovian
case at different values of the QD-1-electrode coupling $\Gamma _{R}$.
(a)-(d) for $\Gamma _{L}/\Gamma _{R}=1$, (e)-(h) for $\Gamma _{L}/\Gamma
_{R}=10$. Here, $C_{k}$ is the zero-frequency ${k}$-order cumulant of
current fluctuations. The non-Markovian effect in
the $\Gamma _{L}/\Gamma _{R}\geq 1$ case has a more remarkable impact on the
first four cumulants of transport current than that in the serially coupled
double QDs system, but the NDC does not appear. The side-coupled double QDs
system parameters: $\epsilon _{1}=\epsilon _{2}=1$, $J=0.001$, $U=5$ and $k_{B}T=0.1$.} \label{fig6}
\end{figure*}

\begin{figure*}[t]
\centerline{\includegraphics[height=12cm,width=16cm]{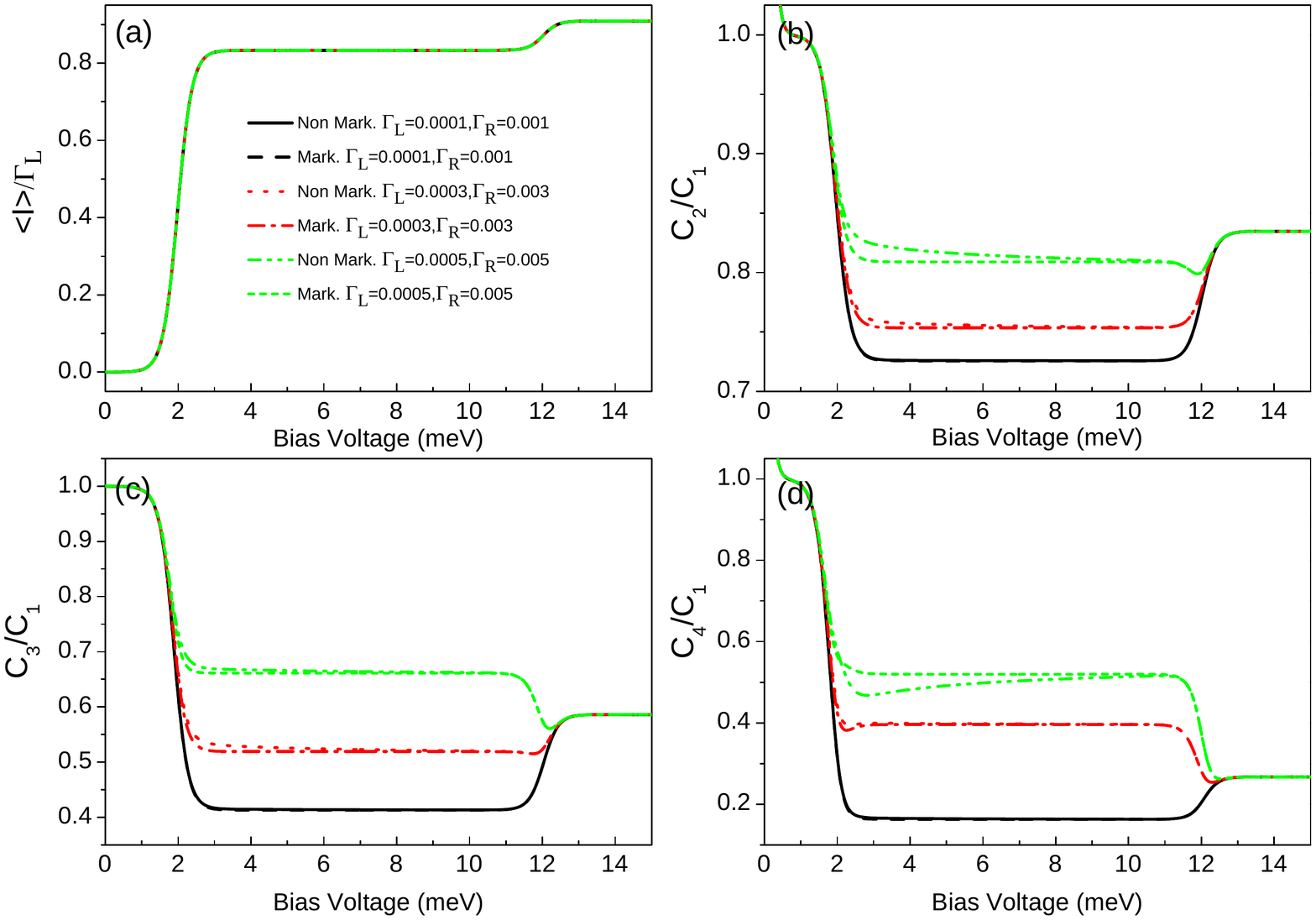}}
\caption{The average current ($\left\langle I\right\rangle$),
shot noise ($C_{2}/C_{1}$), skewness ($C_{3}/C_{1}$) and kurtosis ($%
C_{4}/C_{1}$) versus bias voltage for the Morkovian and the non-Markovian
case at different values of the QD-1-electrode coupling $\Gamma _{R}$ with $%
\Gamma _{L}/\Gamma _{R}=0.1$. Here, $C_{k}$ is the zero-frequency ${k}$%
-order cumulant of current fluctuations. The non-Markovian effect in the $%
\Gamma _{L}/\Gamma _{R}=0.1$ case has a slightly influence on the the first
four current cumulants. The other system parameters are the same as in Fig. 6.} \label{fig7}
\end{figure*}

\end{document}